\documentclass[structabstract]{aa} % for the letters
\usepackage{graphicx} % FJ: Added subfigure
\usepackage{float}
\usepackage{txfonts}
\usepackage{natbib}
\usepackage{units}
\usepackage{url}
\newcommand{\lsp}{LS~I~+61$^{\circ}$303}
\newcommand{\lsi}{LS~I~+61$^{\circ}$303~}
\newcommand{\grs}{GRS 1915+105~}

\newcommand{\beq}{\begin{equation}}
\newcommand{\eneq}{\end{equation}}
\begin{document}

\title{Long-term periodicity in \lsi as  beat frequency between
  orbital and precessional rate}
\author{
M.\ Massi
and F.\ Jaron
}

\institute{
 Max-Planck-Institut f\"ur Radioastronomie, Auf dem H\"ugel 69,
 D-53121 Bonn, Germany \\
\email{mmassi,  fjaron,  @mpifr-bonn.mpg.de}
}

   \date{Received 2012; }
\abstract
{In the binary system  \lsi the peak flux density of the radio outburst, 
which is  related to the orbital period of $\unit[26.4960 \pm 0.0028]{d}$,
  exibits  a   modulation of 1667$\pm$8~d. 
  The radio emission at high spatial resolution appears
  structured in a precessing jet with a  precessional period of  27$-$28~d.}
 {How close is the precessional period of the radio jet to the orbital
  period? Any periodicity in the radio emission should be revealed by
   timing analysis. The aim of this work is to establish the accurate
  value of the precessional period.}
{We analyzed 6.7~years of the  Green Bank
  Interferometer  database at 2.2\,GHz and 8.3\,GHz with the Lomb-Scargle
  and phase dispersion minimization (PDM) methods and performed
  simulations.}
{The  periodograms show two 
	periodicities, $P_1 = \unit[26.49 \pm 0.07]{d}$ ($\nu_1=\unit[0.03775]{d^{-1}}$) and 	
	$P_2 = \unit[26.92 \pm 0.07]{d}$ ($\nu_2 = \unit[0.03715]{d^{-1}}$).	
Whereas   radio outbursts have been known  to have nearly orbital occurrence 
$P_1$ with timing residuals exhibiting a puzzling sawtooth pattern,   
 we probe in this paper that they  are actually periodical outbursts and that their period
is  
$P_{\rm average}=  {2\over{\nu_1  + \nu_2}}=  \unit[26.70 \pm 0.05]{d}$.
The period  $P_{\rm average}$ as well as    
 the  long-term modulation $P_{\rm beat}=     {1\over{\nu_1  - \nu_2}}= \unit[1667 \pm 393]{d}$
result from the 
beat  of the two close periods, the orbital $P_1$ and the precessional $P_2$ periods.}
{The precessional period, indicated by the astrometry to be  of 27--28 d, is $P_2= \unit[26.92]{d}$.
The system \lsi seems to  be  one more case in astronomy of beat, i.e., a phenomenon occurring when two physical 
processes create stable variations  of  nearly equal frequencies.  The very small difference in frequency
creates a long-term variation of period 1/($\nu_1-\nu_2$).
The long-term modulation of 1667 d  results from the beat of the two close
orbital and precessional rates. 
} 
\titlerunning{Orbital and precessional periodicities in \lsp}
\maketitle
\section{Introduction}

The TeV  emitting source \lsi has 
radio characteristics that  make it  unique not only among 
the small number of gamma-ray emitting systems given in  the X-ray binary  class, 
a class of binary systems where a neutron star or a black hole is  orbiting around a  normal  star,
but also  among the larger group of the  
radio emitting X-ray binary systems \citep{fender97, massi05, mirabel12}.
The peak flux density of the radio outburst,   
which is  related to the orbital period of $\unit[26.4960 \pm 0.0028]{d}$,
 exibits  a   modulation of 1667$\pm$8~d
\citep{gregory02}. 
Some double peaked outbursts when observed at two frequencies show 
 different spectral characteristics.
There is a first outburst with a 
flat/inverted spectrum and  a  second  
optically thin  outburst associated with 
different conditions, as indicated by  its  high amplitude, the spectral index, 
and the H$\alpha$ emission line measurements 
\citep{massikaufman09, grundstrom07}.
The complex spectral sequence found in \lsi finds a natural explanation
in the disk-jet coupling model for microquasars: first, there is a continuous
outflow with a flat or inverted spectrum, then an event  triggers a shock in
this slow optically thick outflow \citep{fender04}, and the  growing shock
creates the optically thin outburst \citep{valtaoja92, hannikainen06}.
One of the characteristics that make \lsi unique 
among the other radio emitting X-ray binary systems  is that 
this spectral evolution,  between inverted and  optically thin spectra,
may occur twice during the orbital period \citep{massikaufman09}.
This agrees well  with the \citet{bondihoyle44}  accretion in an eccentric orbit
(as in  \lsp) 
predicting two events along the orbit as shown for \lsi by several authors 
\citep{taylor92, martiparedes95, boschramon06, romero07}.

The binary system \lsp, for which the nature of the compact object 
has  not yet been established
(i.e.,  a black hole or a neutron star),
shares the remarkable property 
of   slow radio quasi-periodic oscillations 
\citep[84 min]{peracaula97}
with the two black hole microquasars  V404 Cyg 
\citep[20--120 min]{hanhjellming92}
and \grs 
\citep[20--40 min]{pooleyfender97, rodriguezmirabel97}.

The radio morphology of the system also shows unique characteristics. 
The resolved  extended structure changes position angle 
(i.e., angle of the projection of the jet onto the plane of the sky, 
measured  from north through east) 
with the surprising large variation
of 60$\degr$  in only one day \citep{massi04, dhawan06}.
Moreover, the jet   is  sometimes
one-sided and at other times two-sided. Because of both of these variations
in position angle and morphology,  
the hypothesis that \lsi might be  a precessing  
microquasar  was brought forth \citep{massi04}.
The one-sidedness of jets is
usually attributed to relativistic bulk motion along a relatively
small angle to the line of sight, which leads to Doppler boosting
of the jet and deboosting of the counterjet emission 
\citep{urrypadovani95}. A variation of that angle due to precession 
would cause variable Doppler deboosting of the counter jet, making it appear for larger
angles (double-sided jet) and disappear for smaller values of the angle 
to the line of sight (one-sided jet, blazar like).

In 2006, VLBA observations by \citet{dhawan06} 
measured the same large rotation of 60${\degr}$/day  
in their images as   \citet{massi04}. 
Some of the VLBA images, showing again a one-sided structure
were, however, interpreted by \citet{dhawan06} 
as  a  cometary tail pointed away from the companion Be star, 
that is in favor of a  pulsar model rather  than of the precessing microquasar 
model.
The reanalysis of this VLBA
data set and the resulting higher dynamic range of the  self-calibrated maps
has   actually revealed    
a double-sided structure in several images \citep{massi12}.  
Before we illustrate  how the results in  \citet{massi12} 
brought us to the present investigation on the precessional period,  
let us consider two important points in the two following paragraphs,  
the first concerning self-calibration and the second 
the pulsar model.

Self-calibration of interferometric data
 is a well-established technique \citep{cornwellfomalont99}.
It may  fail  at  SNR\,$<$\,4   \citep{martividal08} 
or may   create spurious symmetrization for unbalanced 
closure phase triangles (resulting when  
a very displaced telescope is added to the array) \citep{massiaaron91}. 
None of these two cases apply to the used set of only  VLBA data, 
where  double-sided structures 
at 8--16 $\sigma$  are present in 6 out the 12 images.
Concerning  the effects on an image of strong variations of flux density during  observations
   \citep{stewart11},
the source \lsi shows a 
low radio flux density at all orbital phases, apart from  the
maximum of the long-term modulation. In these epochs 
a large outburst lasting few days
occurs around apastron.
This means that  during the
maximum of the long-term modulation one might  expect 
a reduction of  the dynamic range of the produced maps around apastron,
i.e., weak features will be lost.
%%Anyway, 
This is not the case for the VLBA
observations of \citet{dhawan06} performed
toward the minimum of the long-term
modulation.

We are therefore faced in \lsi with a changing structure from
a double-sided structure   to a  one-sided structure.
Are such variations possible also in the pulsar scenario? 
Simulations
\citep{moldon12}  show that the emission from the cometary tail of a pulsar 
for a particular orientation and inclination of the orbit, 
after almost one orbital cycle and at a particular orbital phase  may 
look like a double-sided nebula at a fixed position angle. 
In these conditions, 
the fading and expanding last part of the  cometary tail  
may appear detached from  the  brightest part of the cometary tail which is closer 
to the orbit. 
This possibility is clearly ruled out for \lsp, where along one orbital cycle
the VLBA images show a  double-sided jet at several different orbital phases 
and at different position angles \citep{massi12}.

We return thus to the scenario   of a  precessing microquasar,  where   
the two-sided jet suffers from variable Doppler boosting because the precession  
continuously changes the angle between  jet
and line of sight \citep{massi04, massi12}.
Deriving the precessional period from the available radio images  
is not straightforward, because  the images   reflect  the variation
of the projected angle on the sky plane and, therefore, a combination 
of the ejection angle, inclination, and angle of the precession cone.
The astrometry provides less biased results
\citep{dhawan06, massi12}. 
The astrometry of the VLBA  observations indicates that the peak in consecutive images describes a
well-defined ellipse, six to seven times larger than the orbit, 
with a period of about 27-28 d \citep{massi12}.
Assuming for the microquasar in \lsi the same core-shift effect typically observed in blazars,
the peak  of each image would then correspond to that  
 part of the jet  where the emission becomes optically thick 
at the observing frequency \citep{kovalev08}. 
Based on this assumption, \citet{massi12} interpreted the  ellipse 
as the possible cross-section of the precession cone of the jet 
at the distance  where the emission at  8.4 GHz  becomes optically thick.  
The determined time span  of  27-28 d  to complete the ellipse is 
a first estimate of the precession period. 

The most likely cause for precession of an accretion disk of a compact object is
an assymetric supernova explosion of the progenitor.
As a result the compact object could be tilted \citep{fragile07}.
In this case either the accretion disk is  coplanar
with the compact object and, therefore, subject to the gravitational torque of the 
Be star or,
instead, the accretion disk is  coplanar with the orbit but
tilted with respect to the compact object which induces,  in the context of general 
relativity,
 Lense-Thirring precession if the
compact object  rotates \citep{massizimmermann10}. A deep investigation
of these or other mechanisms of precession requires the knowledge of the precession parameters, such as the period of precession
and the angle of the precession cone.

In this paper we present
a timing analysis of 6.7~years of Green Bank Interferometer (GBI) radio data
aimed at a more accurate determination of the precession period.
Sections  2.1 and 2.2 present the determined  period, called  $P_2$. 
The section illustrates  the case that, while the aim of our research was reached,
i.e., we obtained  a more precise value of $P_2$, 
we were presented with an additional 
unexpected result:  
the beating between the orbital period $P_1$  and the precessional period $P_2$
gives rise to a  new period, $P_{\rm average}=2/(\nu_1+\nu_2)$ modulated
by $1/(\nu_1-\nu_2)=1667$\,d, i.e., the long-term modulation. 
Is the found  $P_{\rm average}$ the periodicity of the
observed radio outburst? 
Indeed, whereas  in the literature $P_1$ is generally referred to as the period
of the radio outbursts, it is also well known  that   
there are differences between  the observed and predicted (for $P=P_1$) outburst times,   
and that  these  timing residuals have vs time a  
 sawtooth pattern (Sect. 2.3).
%%%%% \citep{gregory99} (Sect. 2.3).
In Sect. 2.4 we present two important results.
First, we demonstrate mathematically that
indeed the sawtooth  function adjusts    $P_1$   to   
$P_{\rm average}$. 
Then we show that  
the GBI data folded with $P_{\rm average}$ present 
an offset of 13 d
at  the  minimum of the long-term modulation
equal to the  sawtooth function   and equal to 
that  predicted by the beat between $P_1$ and $P_2$.
In the same section we also present the  corresponding physical scenario.
In Sect. 3 we discuss the  implications of our results
for the observed periodicity in the equivalent width of the $H\alpha$ emission line in \lsp.
In Sect. 4 we present our conclusions.

\begin{figure*}[]
   \centering
   \includegraphics[scale=.6, clip]{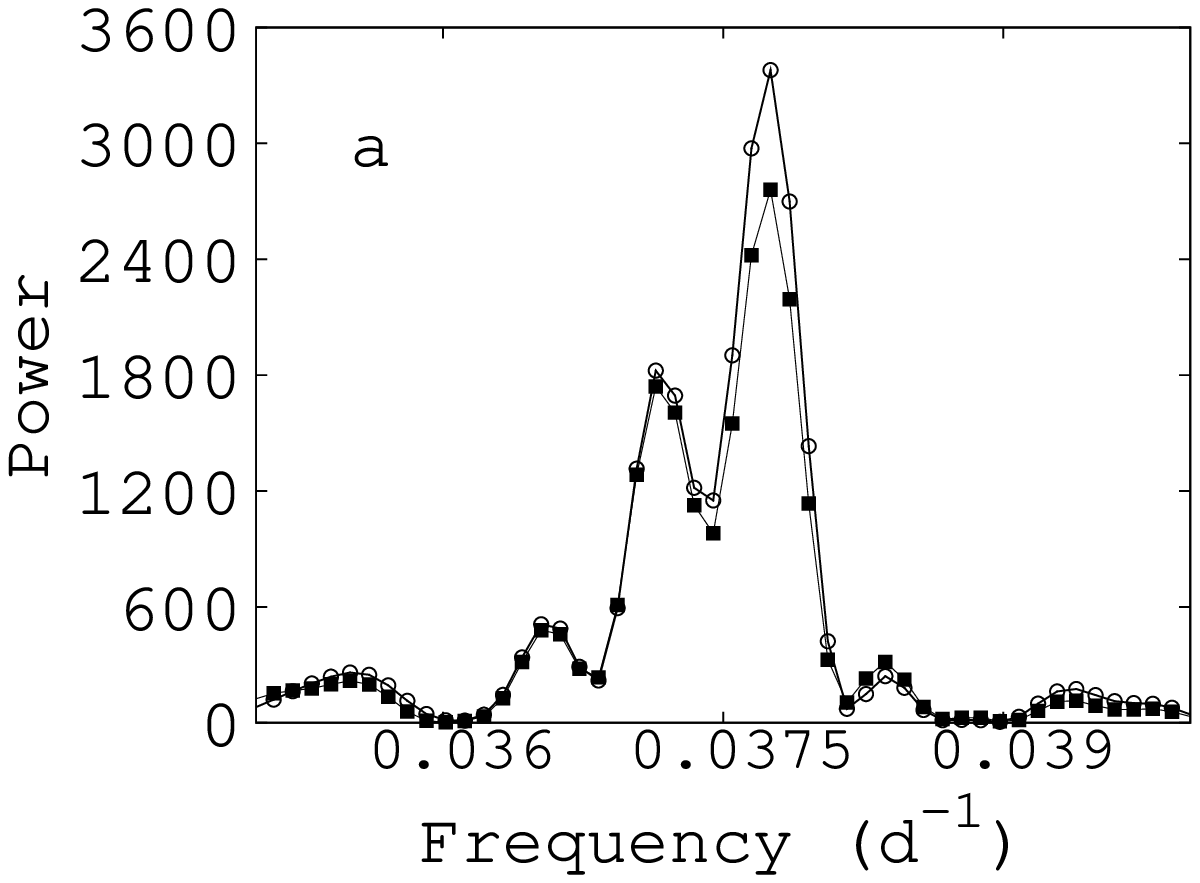}
   \includegraphics[scale=.6, clip]{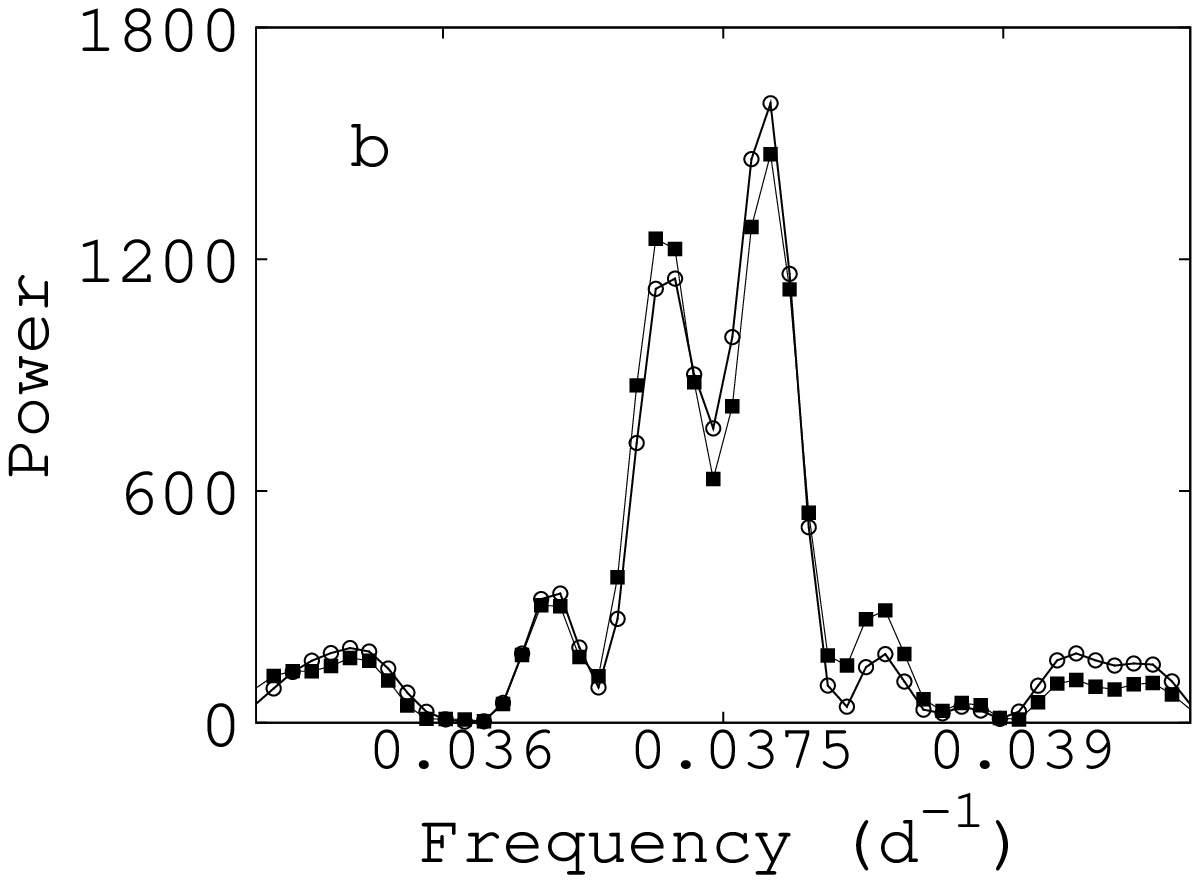}\\
   \includegraphics[scale=.6, clip]{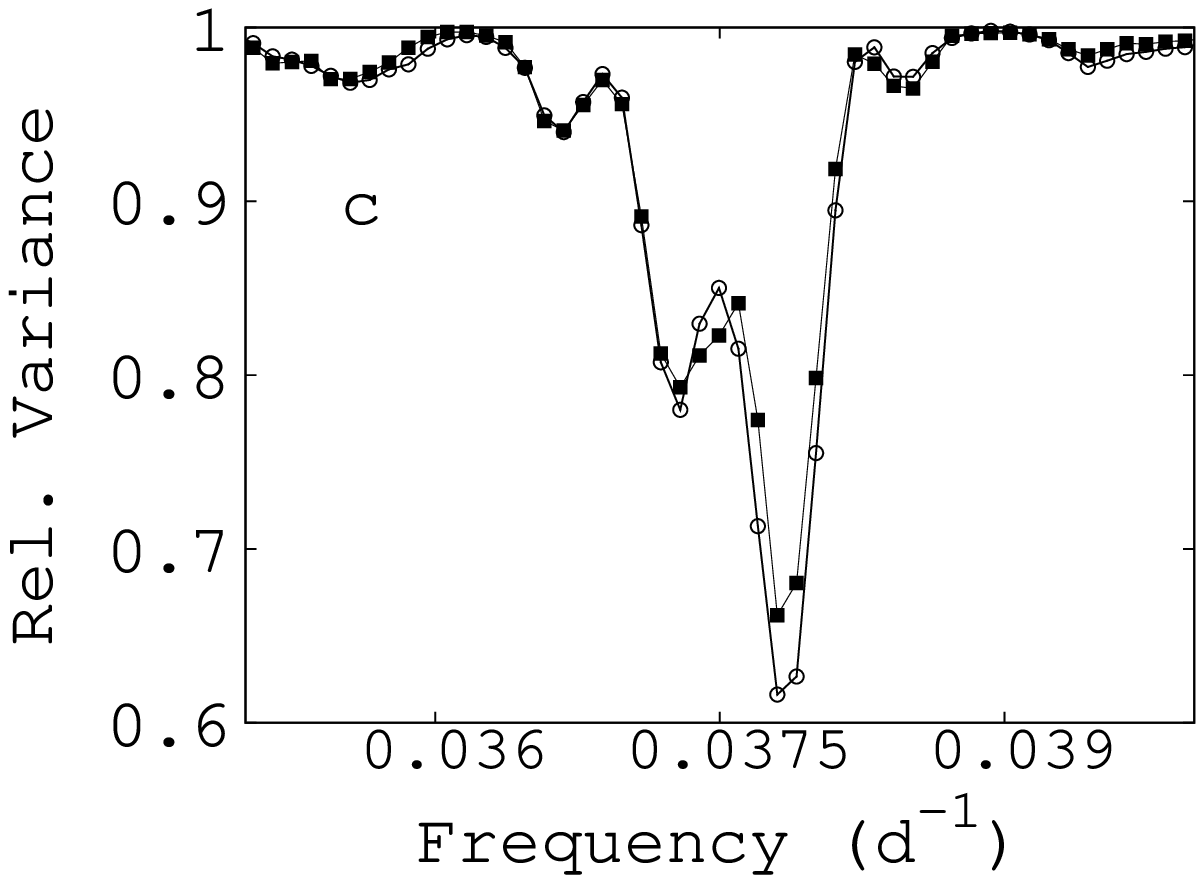}
   \includegraphics[scale=.6, clip]{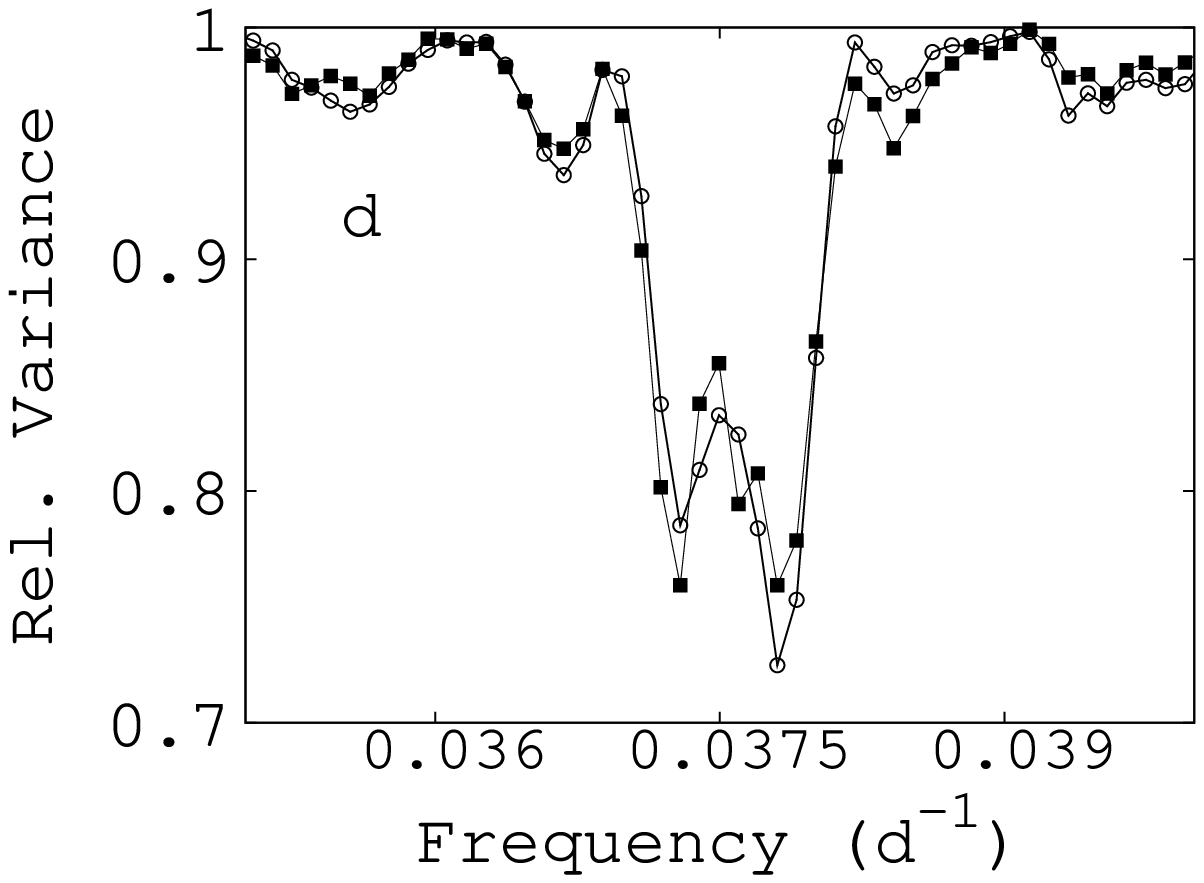}\\
   \caption{Periodograms of 8.3\,GHz (filled squares) and 2.2\,GHz
     (circles)
     data.  a: output of Lomb-Scargle method.  The Lomb-Scargle
     analysis gives on the y-axis the significance of the
     frequency. Two frequencies are found at
     $\nu_1=\unit[0.03775]{d^{-1}}$ ($P_1 = \unit[26.49]{d}$) and at 
     $\nu_2 = \unit[0.03715]{d^{-1}}$ ($P_2 = \unit[26.92]{d}$). 
     b: output of Lomb-Scargle method for data $\ge 4\sigma$. 
     c: output of PDM. The most likely period yields the minimum
     dispersion and appears as a
     minimum in the PDM curve, i.e., specular to the maximum in the
     Lomb-Scargle plot. 
      d: output of PDM for data $\ge
     4\sigma$.}
   \label{Fig1}
\end{figure*}
\section{Data analysis and results}
We  analyzed here  6.7~years of the  NASA/NRAO GBI
\lsi database at 2.2\,GHz and 8.3\,GHz. The database covers three
periods: 49379.975$-$50174.710~MJD, 50410.044$-$51664.879~MJD, and
51798.333$-$51823.441~MJD. The samples, flux densities at both
frequencies with their corresponding errors, in each of the three time
intervals are almost continuous with an average of eight observations per day. In
order to search for possible periodicities we used the Lomb-Scargle
method, which is very  efficient  on irregularly sampled data
\citep{lomb76, scargle82}. We used the algorithms of the UK Starlink software
package, PERIOD (\url{http://www.starlink.rl.ac.uk/}).
For both data sets at 2.2~GHz and 8.3~GHz (Fig.~1) we obtained the same result:
two periods $P_1 = \unit[26.49 \pm 0.07]{d}$ ($\nu_1=\unit[0.03775 \pm  0.00010]{d^{-1}}$)
and $P_2 = \unit[26.92 \pm 0.07]{d}$ ($\nu_2 = \unit[0.03715 \pm  0.00010]{d^{-1}}$).
The statistical significance of a period is calculated in PERIOD
following the method of Fisher randomization as outlined in
\citet{nemec85}. The advantage of using a Monte-Carlo- or
randomization-test is that it is distribution-free and that it is not
constrained by any specific noise models (Poisson, Gaussian, etc.). The
fundamental assumption is that if there is no periodic signal in the time
series data, then the measured values are independent of their
observation times and are likely to have occurred in any other
order. One thousand randomized time series are formed and the
periodograms calculated. 
The proportion of permutations that give a peak power higher than that
of the original time series would then provide an estimate of $p$, the
probability that, for a given frequency window there is no periodic component
present in the data with this period. A derived period is defined as
significant for $p < 0.01$, and a marginally significant one for $0.01
< p < 0.10$ \citep{nemec85}. 
 For both  periods $P_1 = \unit[26.49 \pm
  0.07]{d}$ (frequency window 0.0374$-$0.0379\,d$^{-1}$) and $P_2 = \unit[26.92
  \pm 0.07]{d}$ (frequency window 0.0369$-$0.0374\,d$^{-1}$) and for both data
sets at 8.3\,GHz and 2.2\,GHz we obtained  $0.00 < p < 0.01$.

\subsection{Relative importance between the two periods and previous
  observations}

Figure~1\,a shows the results of the Lomb-Scargle analysis for
the data at 8.3\,GHz and 2.2\,GHz. There $P_1$  is dominating over $P_2$
for a factor of 1.8 at 2.2\, GHz, and a factor of 1.5 at
8.3\,GHz. Figure~1\,b shows the results of the
Lomb-Scargle analysis, if  only data with flux density $\geq 4\sigma$
are used. In this case the two periods have a more comparable
significance, i.e., there is a factor of 1.4 at 2.2\, GHz, and a factor
of 1.2 at 8.3\,GHz. 
We compare the
Lomb-Scargle results with those obtained with 
the phase dispersion
minimization (PDM) method \citep{stellingwerf78}. The results of the
PDM analysis on the whole data sets are shown in  Fig.~1\,c;
the results on data $\geq 4 \sigma$ are shown in Fig.~1\,d. The
results of the PDM analysis agree very well with those of
Lomb-Scargle: there is a different significance of the two periods
when data with low signal-to-noise-ratio (snr) are present 
in the analyzed data set,
i.e., $P_1$ dominates. 

This could be the explanation why in the past
the second period $P_2$ was unseen. \citet{taylorgregory82} found a
period of $\unit[26.52 \pm 0.04]{d}$ in their radio data set; in
1984 they wrote that part of the previous measurements were taken with 
the source in a weak state and repeated the analysis using new data
and also including the old ones, obtaining the value of $\unit[26.496
  \pm 0.008]{d}$ \citep{taylorgregory84}. In 1997 Ray et al.\ reported
new observations and gave a period of $\unit[26.69 \pm 0.02]{d}$,
i.e., coincident with 
our average 
$P_{\rm average}= 
\frac{2}{\nu_1+\nu_2} = 
\unit[\frac{2}{0.03775 + 0.03715}]{d} =  \unit[26.70 \pm 0.05]{d}$,
as discussed in Sect. 2.2.
These observations \citep{ray97} are those of our first GBI
interval, i.e., 49379.975$-$50174.710~MJD. In Fig.~2\,a one sees that
these data sample only the interval of maximum activity.
The difference between the results of \citet{taylorgregory84} and of
\citet{ray97} is puzzling for both groups. \citet{ray97} discuss how
their best estimate of the period is significantly different
($9\sigma$) from the $\unit[26.496 \pm 0.08]{d}$ value of
\citet{taylorgregory84}. \citet{gregory99}, faced with the difference
between their value and the \citet{ray97} results, discuss how unlikely a
sudden change in period would be. In the light of our present
result we see that it is not a sudden change in period but the
presence of two periods that in the Ray et al.\ data have  comparable
significance.
As noted above, the value of \citet{ray97}  $\unit[26.69 \pm 0.02]{d}$ corresponds to our 
$P_{\rm average}$, as discussed in the next section. 

\begin{figure*}[]
  \centering

  \includegraphics[scale=.34]{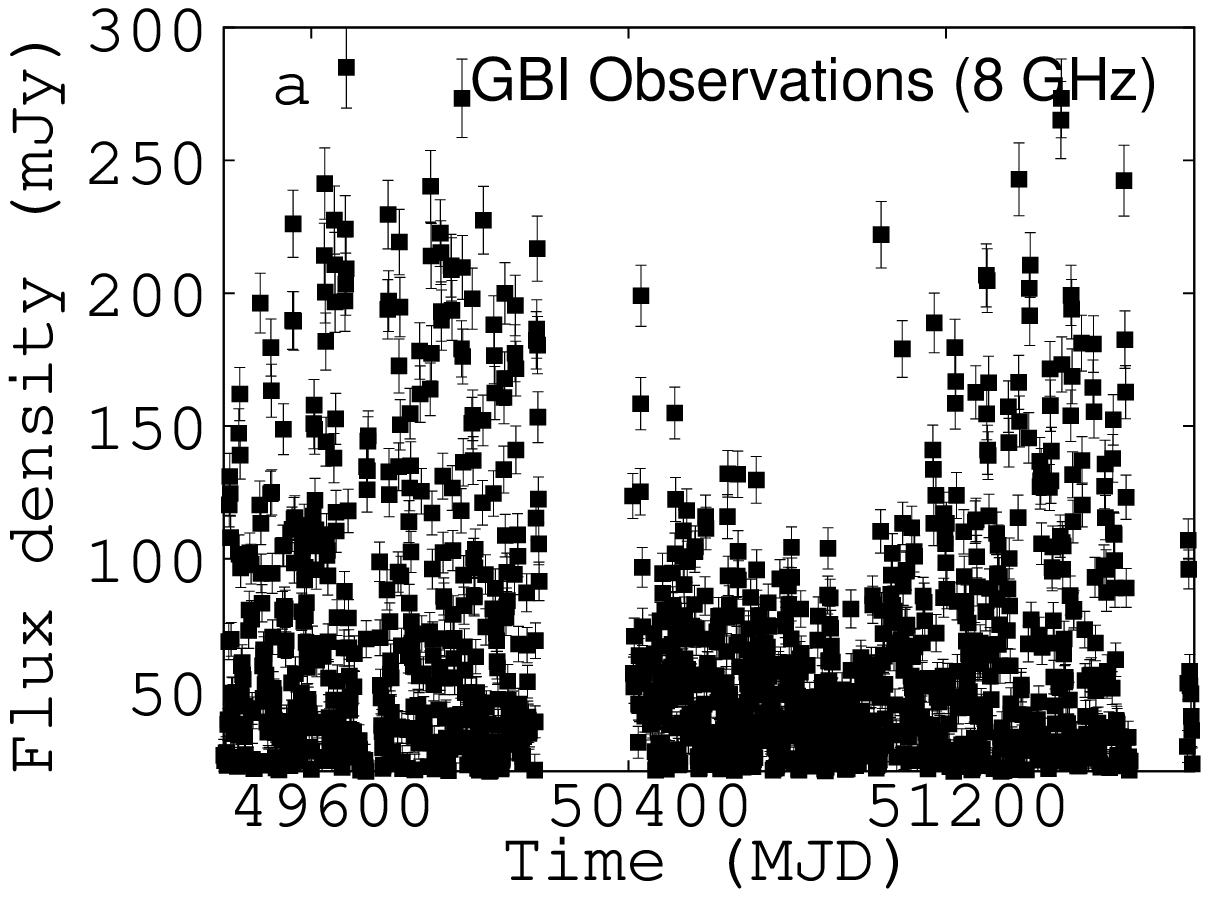}
  \includegraphics[scale=.34]{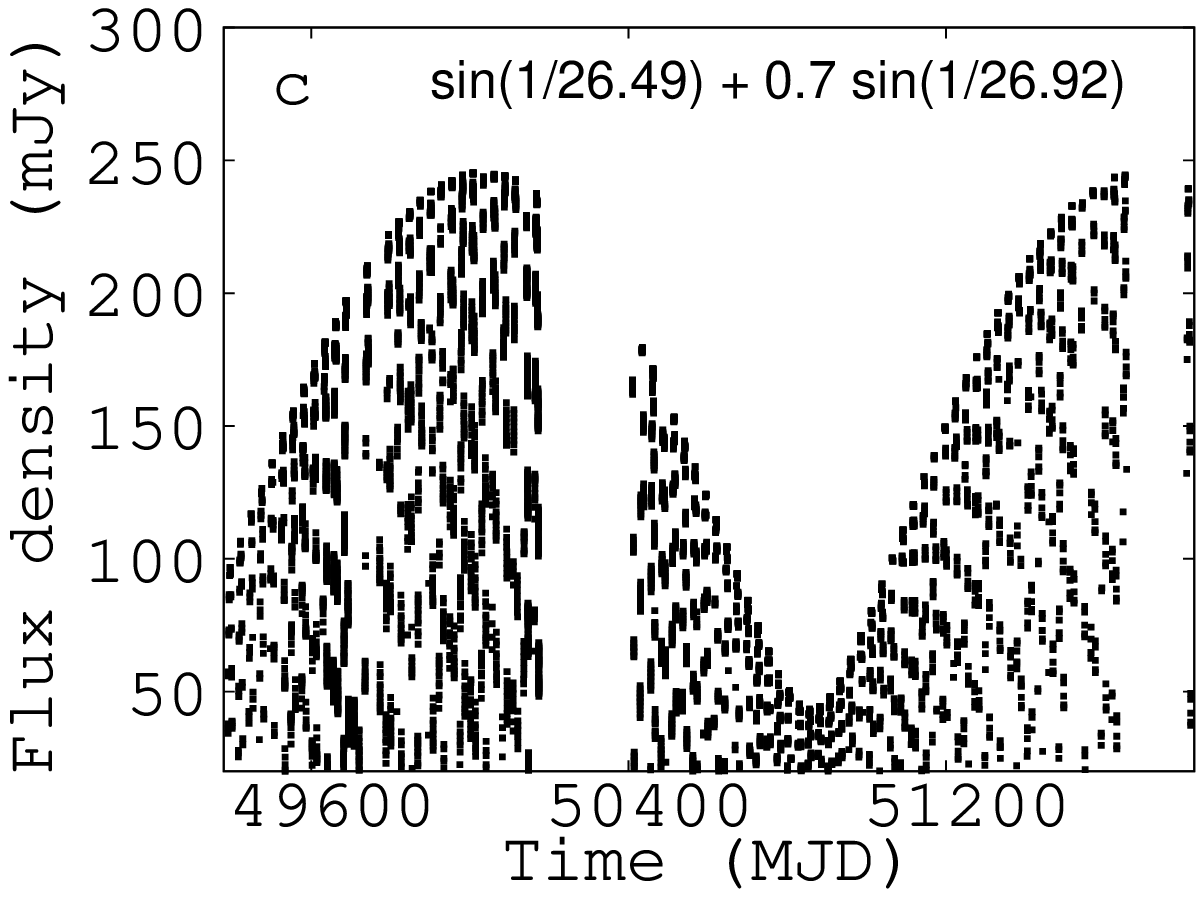}
  \includegraphics[scale=.34]{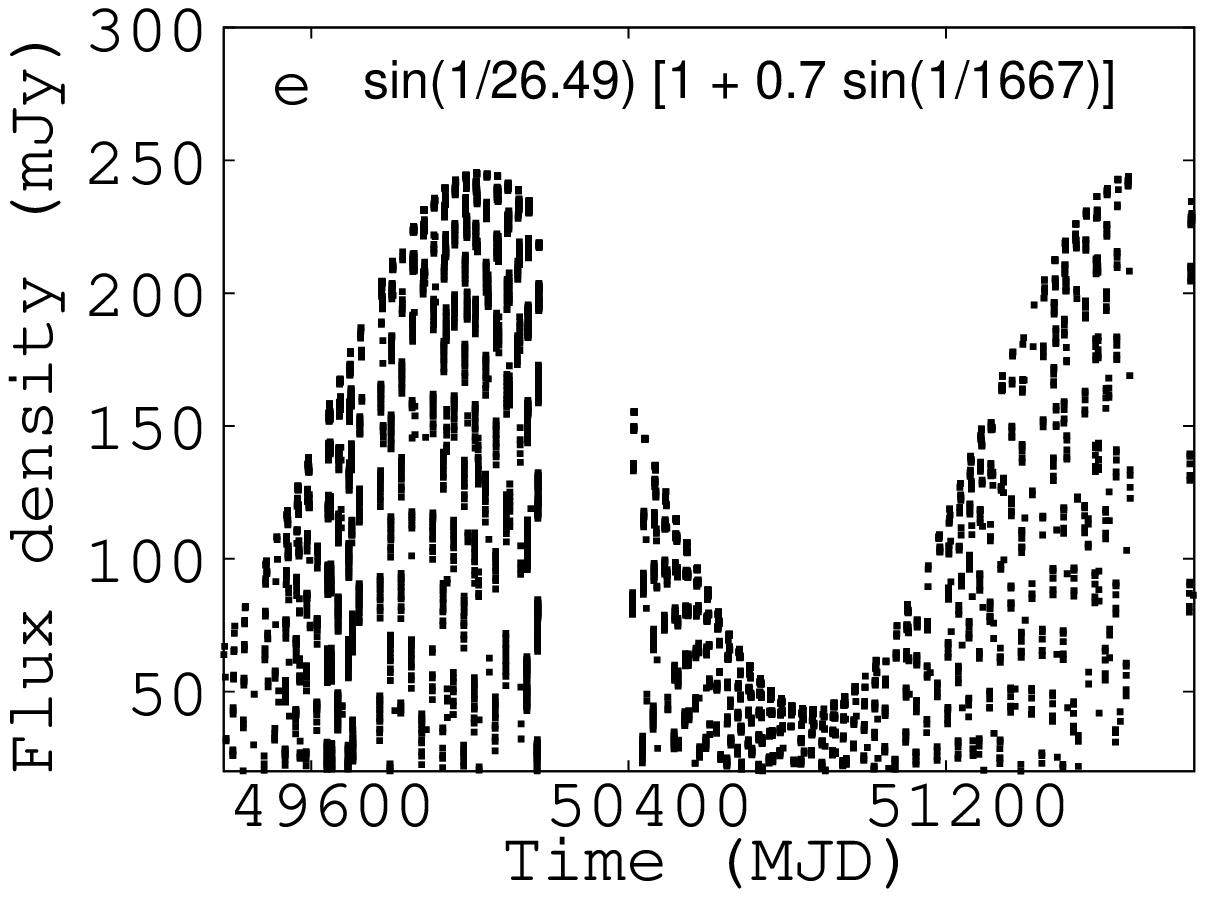}
  \includegraphics[scale=.34]{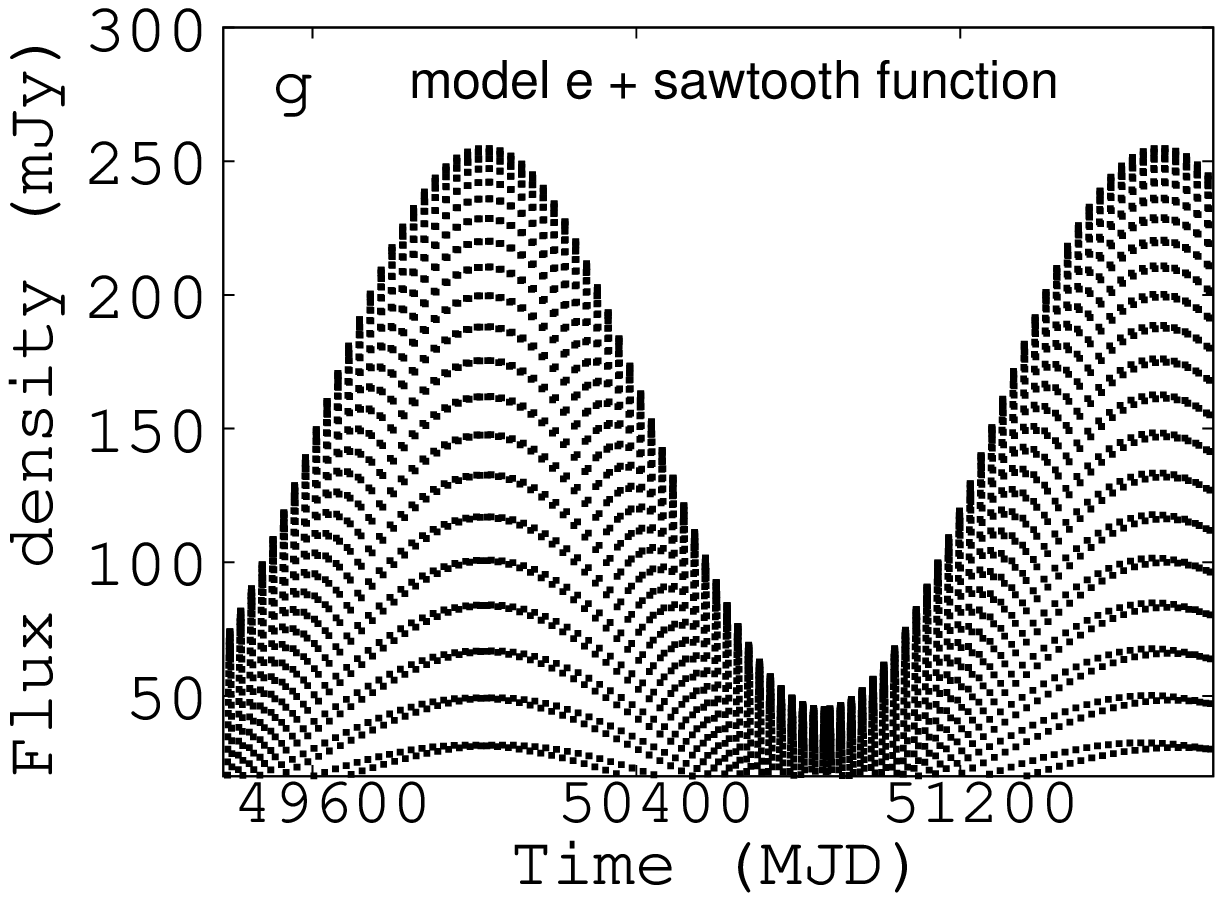}\\
  \includegraphics[scale=.34]{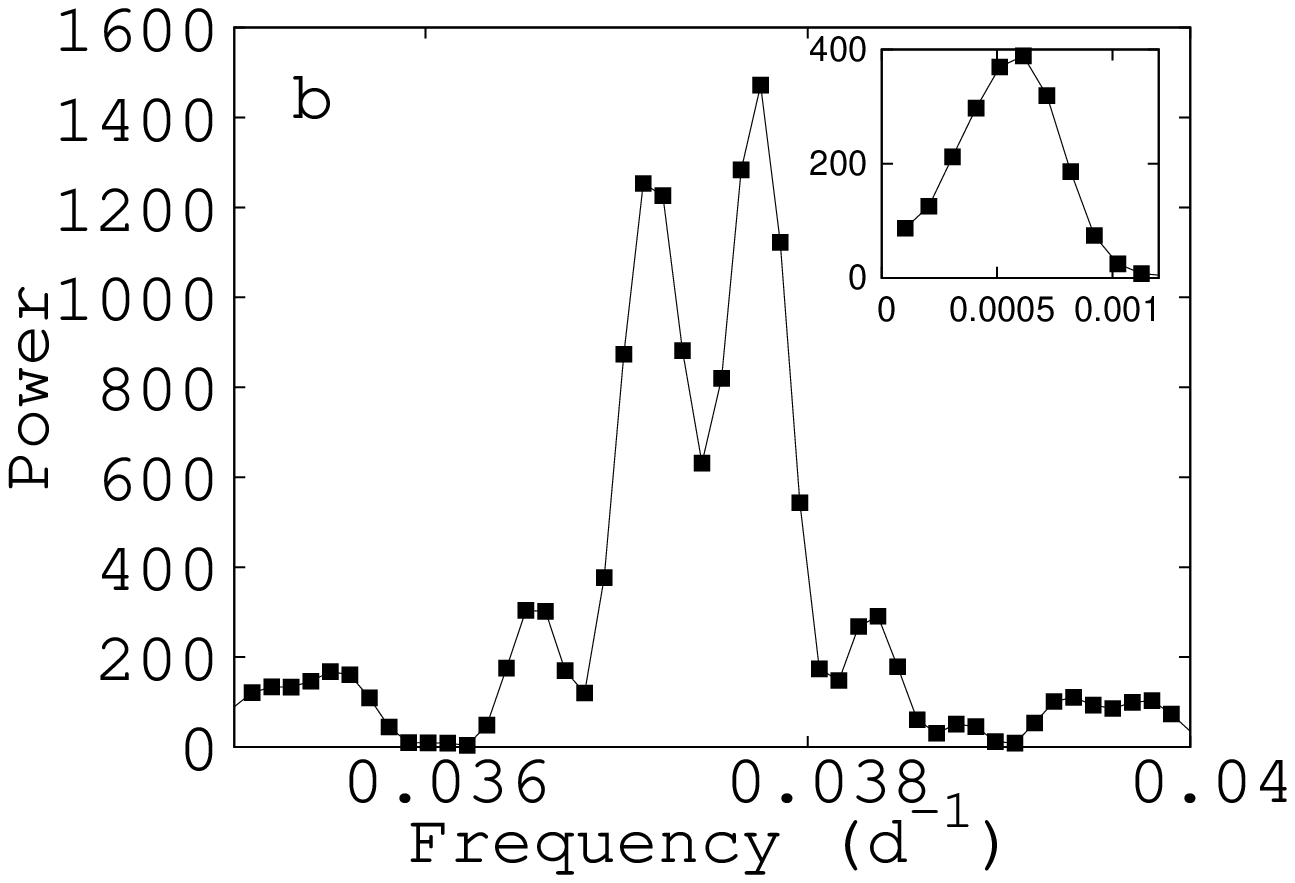}
  \includegraphics[scale=.34]{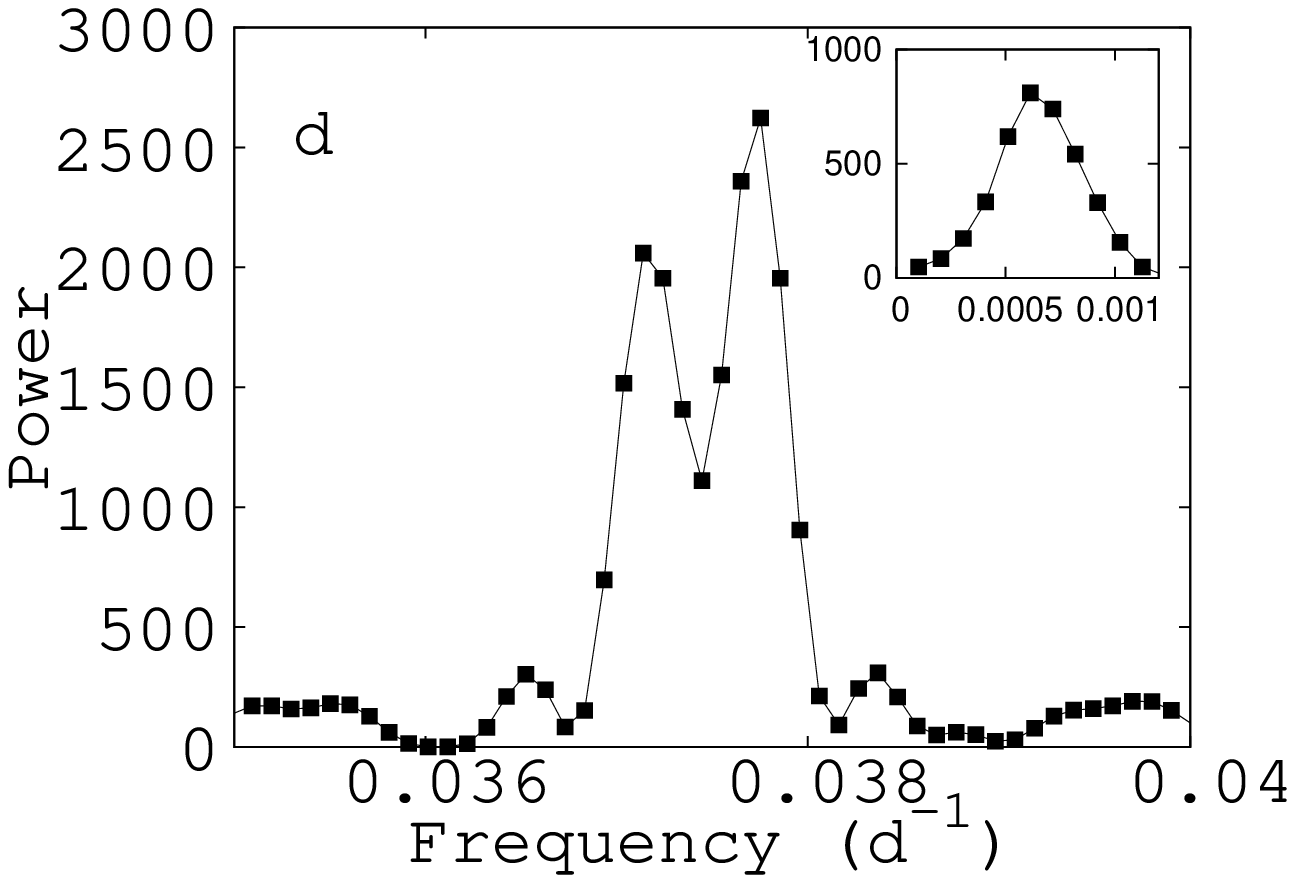}
  \includegraphics[scale=.34]{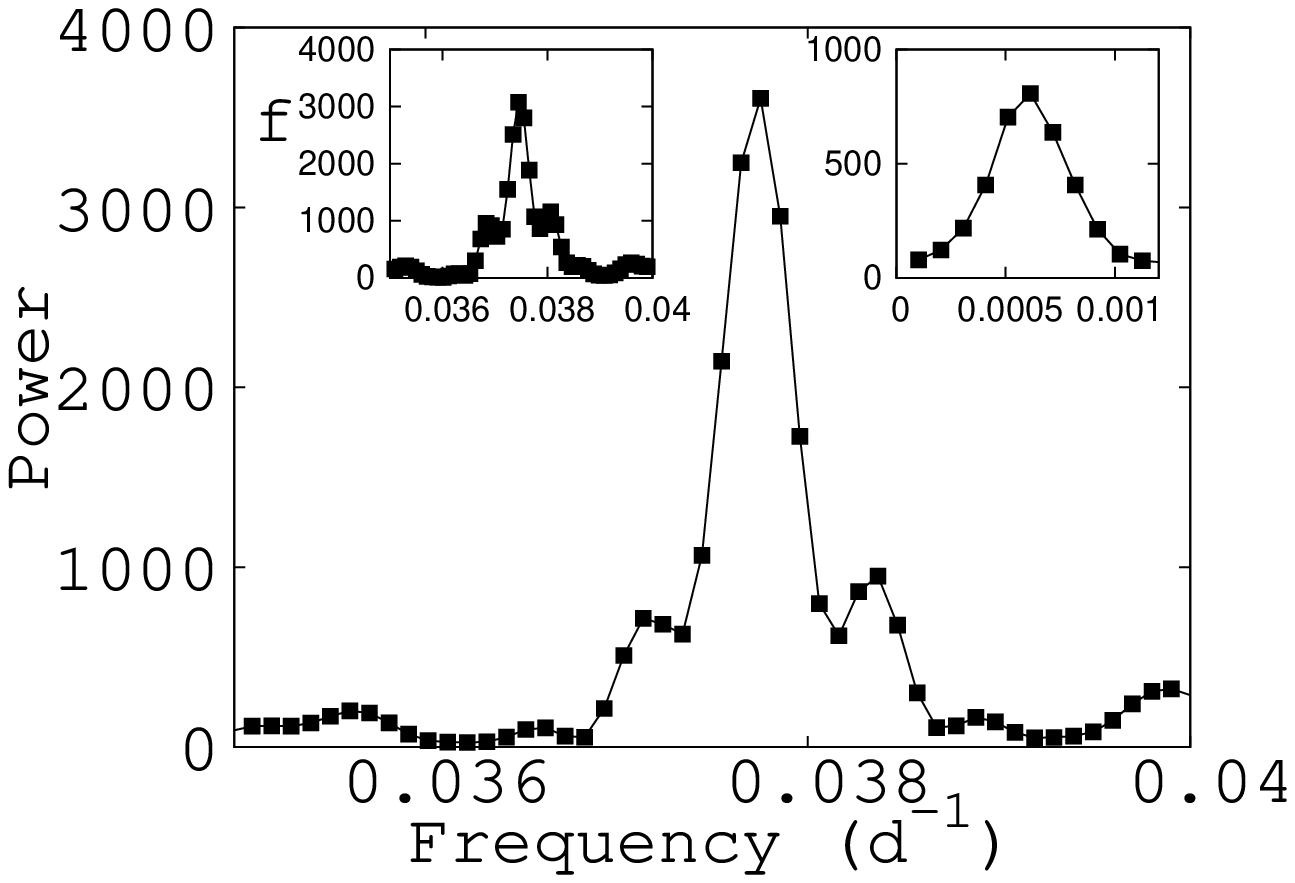}
  \includegraphics[scale=.34]{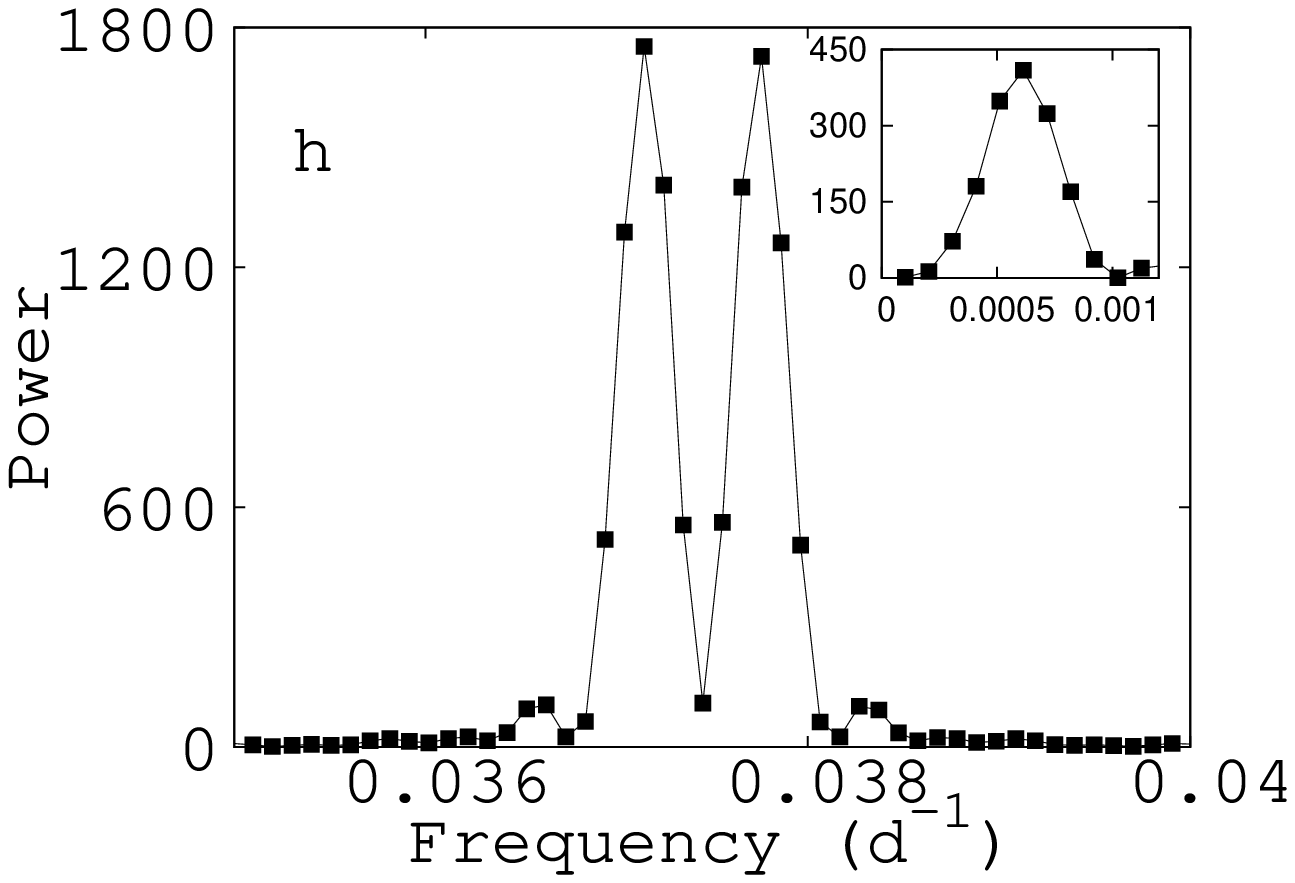}\\

  \caption{Long-term modulation and period analysis. 
    a: 8.3~GHz GBI radio data averaged over 3 d. 
    b: Lomb-Scargle analysis results: two frequencies at
    $\nu_1=\unit[0.03775]{d^{-1}}$ ($P_1=\unit[26.49]{d}$) and
    $\nu_2=\unit[0.03715]{d^{-1}}$ ($P_2=\unit[26.92]{d}$). 
The small window shows the peak at 1/1667~d$^{-1}$ present in all periodograms.
    c: Sum of two sinusoidal functions at 26.49~d and 26.92~d, with an
    amplitude ratio 1/0.7. 
    d: Lomb-Scargle analysis results: two frequencies at
    $\unit[\frac{1}{26.49}]{d^{-1}}$ and
    $\unit[\frac{1}{26.92}]{d^{-1}}$. The significance of the two
    frequencies becomes identical in the periodogram only for an
    amplitude ratio 1/1.
    e: Long-term modulation (1667~d) of a 26.49\,d periodic outburst.
    f: Lomb-Scargle analysis results: one frequency at ${1\over
    26.49}{\rm d}^{-1}$. 
The small window to the left shows the peak at $P_{\rm average}$
present in the periodogram of a simulation of  long-term modulation (1667~d) of a 26.70\,d periodic outburst. 
    g: Sine wave of periodicity $P_1$, modulated
    by a sine wave of periodicity 1667~d and corrected by a sawtooth
    function. 
    h:  Lomb-Scargle analysis results: two frequencies at
	    $\unit[\frac{1}{26.49}]{d^{-1}}$ and
	        $\unit[\frac{1}{26.92}]{d^{-1}}$ as in Figs.~2\,b and d.}
  \label{Fig2}
\end{figure*}
\subsection{Beating: long-term modulation and $P_{\rm average}$}

The two frequencies  $\nu_1=\unit[0.03775]{d^{-1}}$ ($P_1=\unit[26.49]{d}$) and
$\nu_2=\unit[0.03715]{d^{-1}}$ ($P_2=\unit[26.92]{d}$)
are only slightly
different. 
This produces a beating, i.e., a new frequency is formed 
$ \nu_{\rm average}={{\nu_1+\nu_2}\over 2}$, modulated with 
$\nu_{\rm beat}=\nu_1-\nu_2$.
For the sum of two sine functions the following identity holds
\begin{equation}
  \label{eq:beat1}
  \sin\left(2\pi \nu_1 t\right) + \sin\left(2\pi
  \nu_2 t\right)
   =  2\cos \left(2\pi{{\nu_1-\nu_2}\over 2} t\right)\sin\left(2\pi
  {{\nu_1+\nu_2}\over 2}t\right),
\end{equation}
where  the beat frequency (or frequency of the envelope) 
$\nu_{\rm beat}=\nu_1-\nu_2$ is twice the frequency of the cosine
term. 
 In our case, the term  ${1\over \nu_1-\nu_2}=\unit[\frac{1}{0.03775 - 0.03715}]{d}$
 is  equal to  $\unit[1667 \pm 393]{d}$.\footnote{$({1\over P_1} - {1\over P_2})^{-1} = 1658\pm 382$ when using $\geq$ 6 digits}

Figure~2\,c shows the sum of two sine functions with different amplitudes 
\begin{eqnarray}
  f_{\rm b}(t) & = & \sin\left(2\pi \nu_1 t\right) + a\sin\left(2\pi
  \nu_2 t\right)\nonumber\\
  & = & 2a\cos \left(2\pi{{\nu_1-\nu_2}\over 2} t\right)\sin\left(2\pi
  {{\nu_1+\nu_2}\over 2}t\right)\nonumber\\
  \label{eq:beat2}
  & & + \left(1 -a\right)\sin\left(2\pi\nu_1t\right),
\end{eqnarray}
with  $a = 0.7$, whereas
Fig.~2\,e  shows the function
\begin{equation}
  \label{eq:mod}
  f_{\rm m}(t) = (1+b\sin(2\pi\nu_{\rm m}t))\sin(2\pi\nu_1t),
\end{equation}
with  $b = 0.7$,
$\nu_1 = \unit[\frac{1}{26.49}]{d^{-1}}$, and  $\nu_{\rm m} =
\unit[\frac{1}{1667}]{d^{-1}}$.
As one can see, both Eqs. 2 and 3 are able to reproduce the long-term modulation; however, the
periodograms are rather different. The periodogram of Eq.~2, shown in   Fig.~2\,d  
agrees well with the periodogram of the GBI data of Fig.~2\,b. On the contrary,  
as one can see
in  Fig.~2\,f, in the periodogram of Eq.~3 
only $P_1 = \unit[26.49]{d}$ is present in the frequency range 0.036$-$0.039 d$^{-1}$.
\subsection{Sawtooth function}

\citet{gregory99a} demonstrated the existence of a long-term modulation of the peak outburst flux.
\citet{gregoryneish02}  indicated that the  modulation in radio properties   may stem from periodic ejections
of a shell (density enhancement)
of gas  in the equatorial disk of the Be star.

The long-term  modulation  is  also present  in the  timing residuals of the outbursts,
i.e., the difference between  observed and predicted (for $P=P_1$) outburst time \citep{gregory99}. 
The observational result is that  timing residuals show a surprising sawtooth
pattern, i.e., with a gradual rise from 0 to  about 6 d, but then a rapid fall 
to a large negative value of about -7 d.
The saw tooth function is  shown in Figs.~2 and 8\,a of \citet{gregory99} and here in 
Fig.~3\,a.
In detail the trend is as follows: observed and predicted outburst times  coincide 
at the peak of the long-term modulation (i.e., at the radio maximum)
resulting in a timing residual equal to  $\tau=0$;
the timing residual grows linearly with time and 
at the minimum of the long-term modulation 
reaches a maximum of about 6~d 
where it  sharply switches to about -7~d. 
Also surprising is that the transition $\tau \simeq 6$\,d 
to  $\tau\simeq -7$\,d is not related to 
a strong change in flux density; it
occurs at a time when the amplitude is  low, i.e., at the minimum
of the long-term modulation (Figs.~6\,a and 7\,b of Gregory et al.~1999). 
After this transition, $\tau$ starts to grow linearly with time reaching 
the value $\tau=0$ only after  about 800 d  at the  
new maximum of the long-term modulation. 

We performed a  test by using the sawtooth function to correct the 
model of Eq.\ \ref{eq:mod}
($P_1$ modulated by 1667~d)  and to verify  if  
the corrected model is able to reproduce the 
observed spectrum (i.e., $P_1$ and $P_2$).
First we have to define the sawtooth function. The slope
of the sawtooth pattern  in Fig.~8\,a of \citet{gregory99}, is about
0.008; we generate therefore the
sawtooth function (Fig.~3\,a)  with a period of 1667~d as 
\footnote{fmod is a function implemented in the
math library of C.}
\begin{equation}
\tau(t) = 0.008\,\mathrm {fmod}\,(t, 1667)  
\end{equation}
 and include it in Eq.\ \ref{eq:mod}, which becomes:
\begin{equation}
  \label{eq:mod+st}
  f_{\rm m}(t) = (1 + b\sin(2\pi\nu_{\rm m}t))\sin(2\pi\nu_1(t - \tau(t))).
\end{equation}

To study the  effect of the sawtooth function without
any bias resulting from large holes in the sampling, 
 we performed both simulations with original 
 sampling and with regular sampling. 
We obtained the same results and here we show  those with regular sampling. 

The resulting periodogram of the simulated data 
(Fig.~2\,g) is shown in Fig.~2\,h. 
One sees that  the model with the single
periodicity\ $P_1$, modulated by 1667~d, once corrected by the sawtooth
function, is able to reproduce  the results of our spectral analysis that is the 
two periods  $P_1$ and $P_2$.
In the next section we show that 
when one directly uses  the two found periods $P_1$ and $P_2$,
the sawtooth function is naturally explained.
 
\subsection{Period of the observed outburst and $P_{\rm average}$}
The sawtooth function results from the comparison between observed and  
predicted  (for $P=P_1$) outburst time.
Here we will show first  analytically (Eq. 6) and then with the GBI observations that
the observed outburst occurs  at   $P=P_{\rm average}$, 
i.e., at 1/$\nu_{\rm average}$ of Eq.~1.

Analytically, adjusting  $P_1=\unit[26.49]{d}$
($\nu_1 = \unit[0.03775]{d^{-1}}$) by the given sawtooth
function\ $\tau(t)$  to
\begin{equation}
\nu_1(t - \tau(t)) = \nu_1 t(1 -0.008) = (0.03775\times 0.992)~t =
0.03745~t
\end{equation}
gives as a result    $P_{\rm average} = \unit[\frac{1}{0.03745}]{d}
= \unit[26.70]{d}$  as in Sect. 2.1,
(with variations between 26.65$-$26.70~d for slopes between 0.006$-$0.008).
This is an important result. It implies that the timing residuals 
between predicted (at $P_1$) and observed outbursts are equal to the residuals between
predicted (at $P_1$)  and $P_{\rm average}$. 

We therefore ascertained that the observed outburst has  periodicity  $P_{\rm average}$. 
However, if we use  Eq. 3 and we  simply substitute $P_1$ by  $P_{\rm average}$,
the periodogram fails to reproduce the observed periodogram with  $P_1$ and $P_2$ and 
just shows $P_{\rm average}$  (small window on the left in Fig.~2\,f).
Is  $P_{\rm average}$  only  an apparent  periodicity, just produced by $P_1$ and $P_2$? 

Let us fold the GBI data on the periods  $P_1$, $P_2$, and $P_{\rm average}$
(Figures~4\,a-c).
The data folded with the orbital period $P_1$ show the
broad cluster that is well known in the literature 
(e.g.,  Fig.~2\,c in Massi \& Kaufman 2009), 
where flares above 100 mJy occur from about phase 0.4 to about phase 0.9. 
The broadening is due to the differences between the observed and predicted (by $P_1$)
 outburst time that also causes the sawtooth pattern.
Figure 4\,b, shows for the first time the data folded with the precessional period $P_2$.
The cluster of the large flares  is also evident and superimposed to scattered smaller flares.
If now we fold the data with $P_{\rm average}$ one would expect, since
it is the average of $P_1$ and $P_2$,
a clustering rather similar to that in  Fig.~4\,a and b. The result, shown in Fig.~4\,c,  is completely different.
First of all, there are two clusters. 
Second, each one of the two clusters is not as broad as those
with $P_1$ and $P_2$, i.e., the  clustering is  better.
Where does the  double clustering come from?
We used the color green for data  after  50841 MJD, located in the minimum.
A harmonic might theoretically  give rise to two possible clusters, but in this case  
green and black points had to be present in both clusters.
On the contrary, all data before the minimum, i.e. the black points,
 cluster at one phase and all points after that, i.e., green points, cluster at another phase.
The points before and after  50841 MJD   cluster separately with a shift of  0.5 in phase (or about 13 d).

We have also folded  the simulated 
data of Eq. 2  with  $P_{\rm average}$. In this case the dependency on $P_1$ and $P_2$ is very simple, just two sine functions.
Nevertheless, the same kind of double clustering shown by the GBI data occurs for the simulated data of Eq. 2  as one
can see in the box of Fig.\,4\,c.  
The fact that a simple sum of sine functions in 
$P_1$ and $P_2$ produces  the same jump,  
as the GBI data, 
if folded with $P_{\rm average}$,  
implies either  that this simple mathematical form is true in \lsp, 
or, more likely,  that the jump
is a property of the beating process and its value only depends on
the two periods $P_1$ and $P_2$.  

In mathematical terms, as shown in the Appendix, the beat  of the two sine functions  
$f(P_1)$ and $f(P_2)$ has a phase reset  when 
the delay of  $f(P_2)$ with respect to  $f(P_1)$ becomes larger than $P_2/2$.  
Until that point  $f(P_1)$  preceeds $f(P_2)$ (and $P_{\rm average}$)
giving rise to a positive timing residual.
After that point   $f(P_2)$  (and $P_{\rm average}$)  preceeds $f(P_1)$.
This produces  the jump from a large timing residual to a 
nearly  equally  large, but  now negative,  timing residual  
observed in the sawtooth function
(see Fig.~3\,c and d and the derived saw tooth function in Fig.~3\,b,
in impressing agreement with the observed one of  Fig.~3\,a).

In physical terms  the  jump is illustrated  in Fig.~4\,d. 
The radio maximum of the long-term modulation results  when the ejection,
 periodical at $P=P_1$,
occurs at the smallest  angle with
respect to the line of sight and the  Doppler boosting is largest
\citep{kaufman02}.
Because of the precession $P_2$ the angle of the ejection changes, and  
the  radio minimum  corresponds to an   ejection 
occurring  at the largest  angle with respect to the line of sight.
At this  point  the ejection has travelled half a
 precession cone and  turns  onto the other half of the precession cone,
i.e., from I to II   in Fig.~4\,d.
This causes the phase jump in the data folded with  $P_{\rm average}$ and the
jump in the  saw tooth function of Fig.~3\,a.

Finally, in  Fig.~4 d, we present the data folded with $P_{\rm average}$, where
we fold the data before 50841MJD (black points) using the usual 
$t_0 =43366.275$MJD   \citep{gregory02},
whereas  for the  data after 50841MJD (green points) we use  $t_0 + 13.25$\,d,
correcting for the jump at the minimum.
The better folding of the large flares with $P_{\rm average}$ 
with respect to those with the two real periodicities $P_1$ and $P_2$ is the observational evidence
for the result of  Eq. 6, i.e., the periodicity of the radio outburst is $P_{\rm average}$.

\begin{figure}[htb]
    \centering
  \includegraphics[width=0.2\textwidth, height=0.14\textwidth]{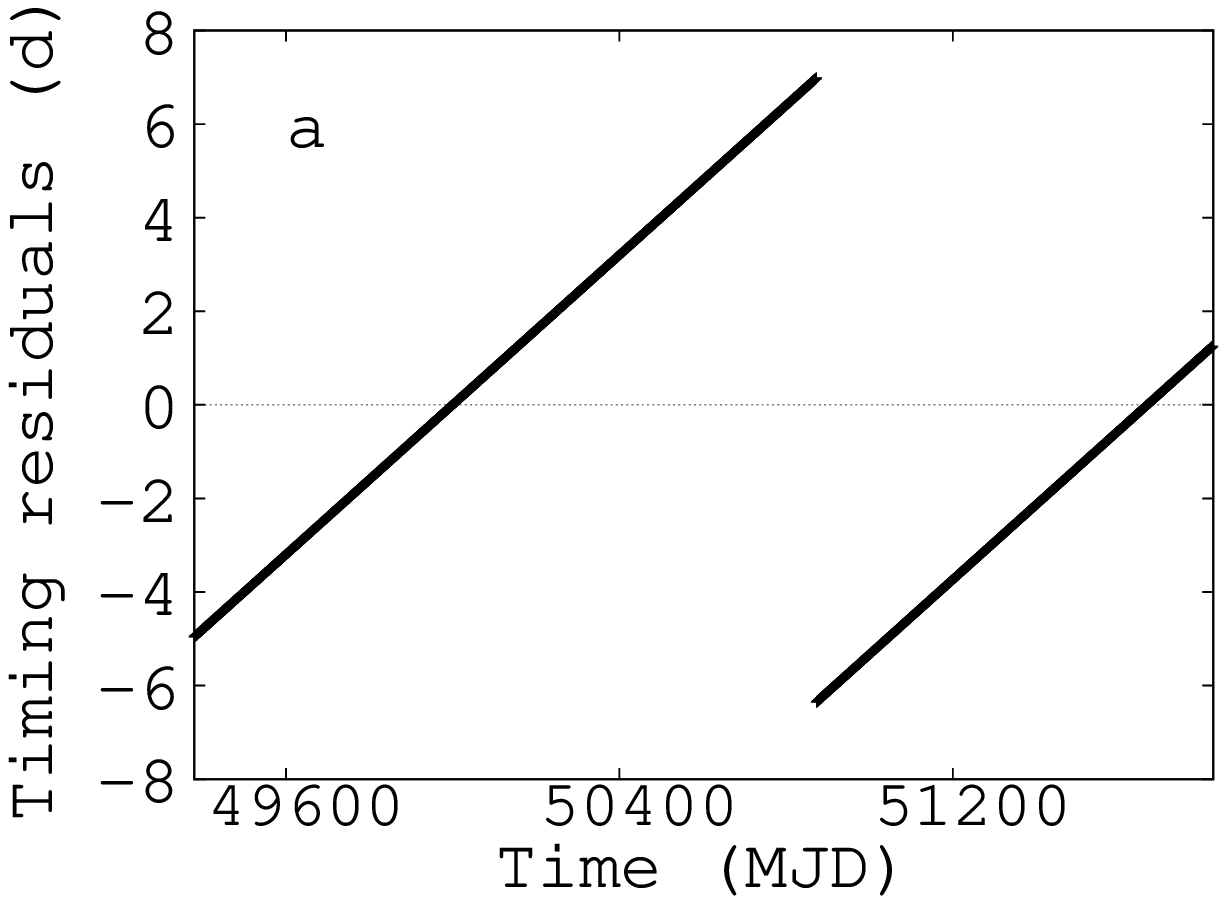}
  \includegraphics[width=0.2\textwidth, height=0.14\textwidth]{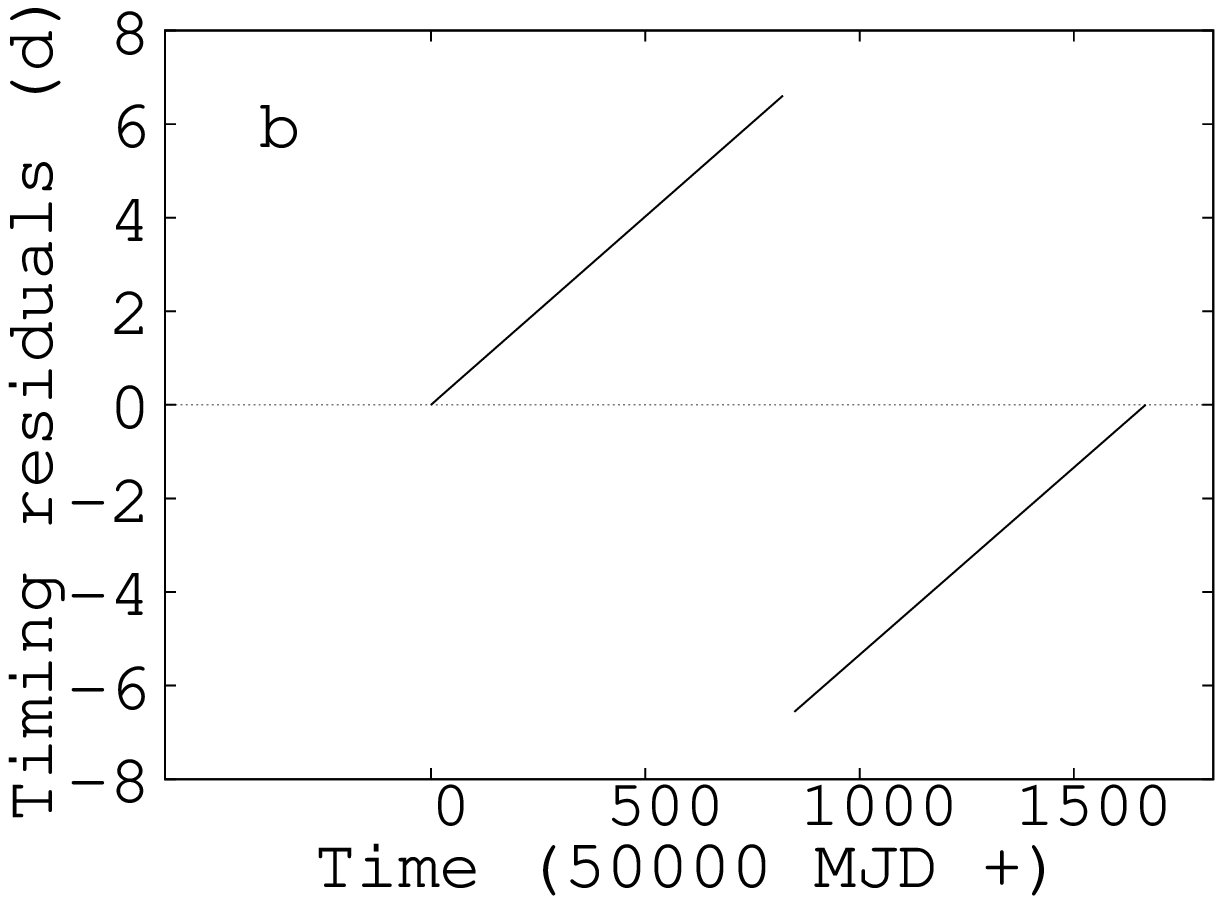}\\
  \includegraphics[width=0.4\textwidth, height=0.14\textwidth]{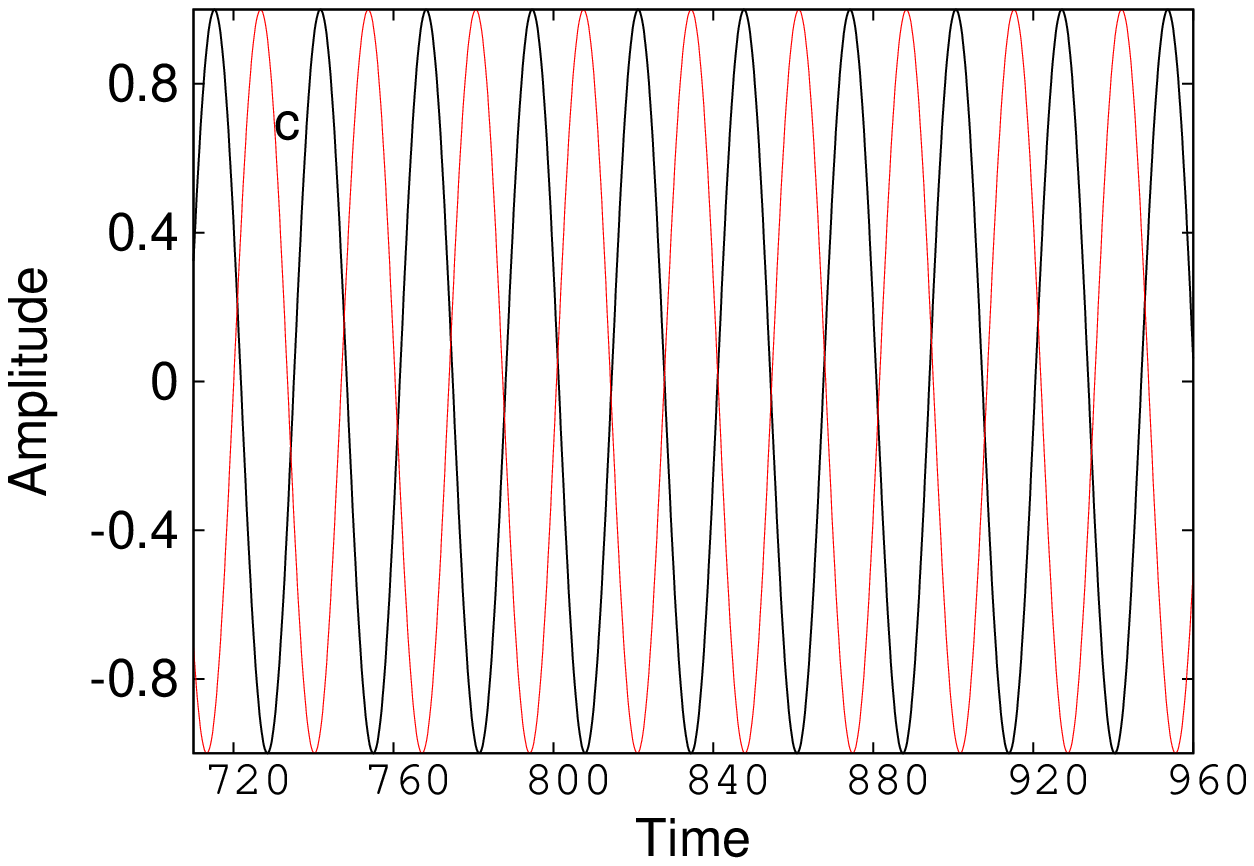}\\
  \includegraphics[width=0.4\textwidth, height=0.14\textwidth]{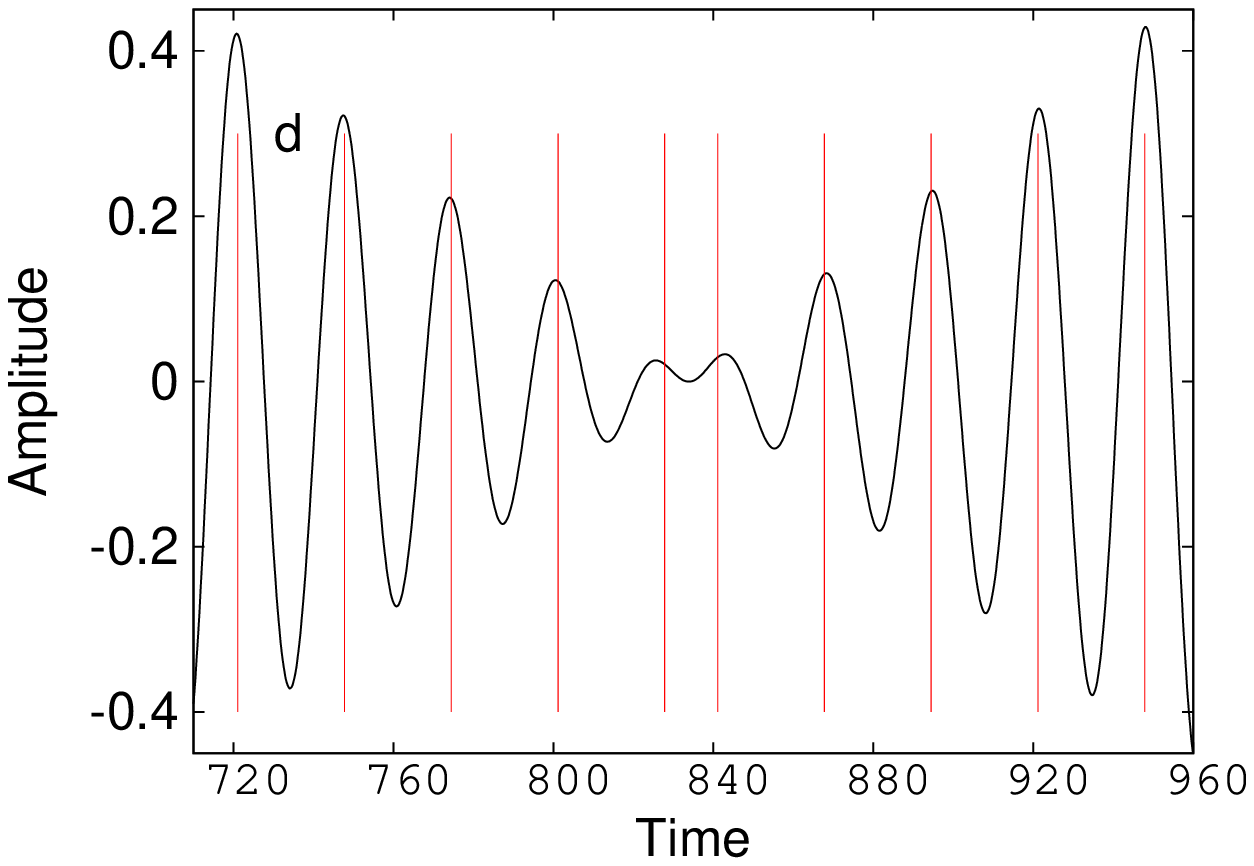}\\
  \caption{ a: Sawtooth function based on timing residuals observed by \citet{gregory99}.
    b: Sawtooth function  based on the loss of synchronization of the sine functions $f(P_1)$ and $f(P_2)$  of Fig.~3\,c.
    c: Sine waves with
     periods $P_1$ (black line) and $P_2$ (red line) starting with synchronized peaks at $t=0$. 
    d: Sum of the sine waves. The first five bars are  at a regular interval
equal to $P_{\rm average}$, then the next bar follows at a distance of
only 13.25 d  ($P_1/2$),
followed by  four bars again  at a regular interval
equal to $P_{\rm average}$ as before. The two central bars  are symmetric
with respect to a  peak of  $f(P_2)$ (Fig.~3\,c) that is nearly equidistant from 
the preceeding peak and the delayed peak of $f(P_1)$ (see Appendix).
}
\end{figure}

\section{Periodicities in the equivalent width of the H$\alpha$ emission line}

Our results have important implications for the short- and long-  H$\alpha$ variations 
for the Be star of the \lsi system.
\citet{zamanov99}  analyzed H$\alpha$ spectra of
\lsi and  determined 
the same 26.5 d radio period, and in addition determined that  
the H$\alpha$ emission line equivalent width (EW) 
varies  over the same time scale 
as the long-term radio modulation.
Moreover,  \citet{zamanov99} tried to find evidence for
long-term periodicities  in other line parameters like 
the important B/R ratio.

Most Be/X-ray binaries show asymmetric split H$\alpha$ profiles,
the ``blue'' (B) or ``violet'' (V) peak and the ``red'' (R) peak. 
B/R (or  V/R) variability refers then to
the variation of the relative strength of the blue to the
red peak. 
B/R  variability cycles are common in Be stars forming canonical Be/X-ray binaries 
containing accreting X-ray pulsars. 
As stated above, Zamanov and collaborators   tried to find evidence in \lsi for
long-term  periodicities  in  the B/R ratio, but they conclude that  
``unfortunately, the results
of this search turned out to be negative'' \citep{zamanov99}.

V/R variability is explained in terms of a nonaxisymmetrical
equatorial disk in which a one-armed
perturbation (a zone in the disk with higher density)
propagates \citep{reig11}.
As a possible explanation  for  the long-term modulation,
\citet{gregoryneish02}  indicated that   it  may stem from periodic ejections
of a shell of gas  in the equatorial disk of the Be star.
\citet{gregoryneish02} commenting  the fact that  there are no
 periodic variations in the H$\alpha$ V/R ratio for \lsp,
nevertheless tried to  test the one-armed density wave model predictions. 
Their conclusion is that the radio behavior  ``is at odds with the predictions of a one-armed density wave model.
 Thus, the one-armed density wave model does not agree quantitatively with the measurements 
for \lsp'' \citep{gregoryneish02}.

In other words, the origin of the long-term variation of  the H$\alpha$ emission line 
from  the disk of the Be star does not  seem to be  related to structural disk variations 
(no B/R variations), whereas it is clearly related to
 a periodical  change in the number of emitted 
photons (the EW(H$\alpha$)).  
A likely candidate as  agent of these variations is  the relativistic precessing jet  
that could well be able to produce 
EW(H$\alpha$) variations 
of the same timescales as those  we observe in  its  radio emission. 

\begin{figure}[]
    \centering
   \includegraphics[width=.22\textwidth, height=.25\textwidth]{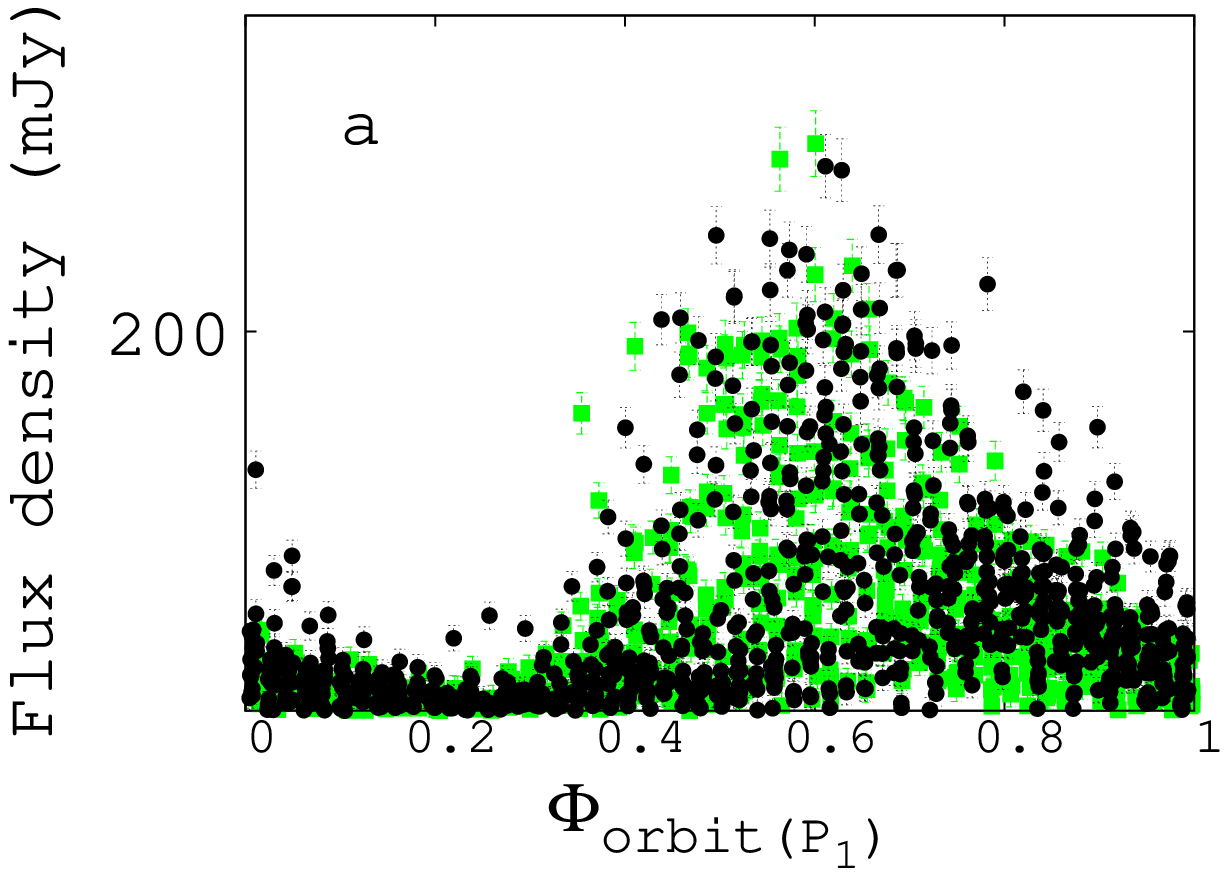}
   \includegraphics[width=.22\textwidth, height=.25\textwidth]{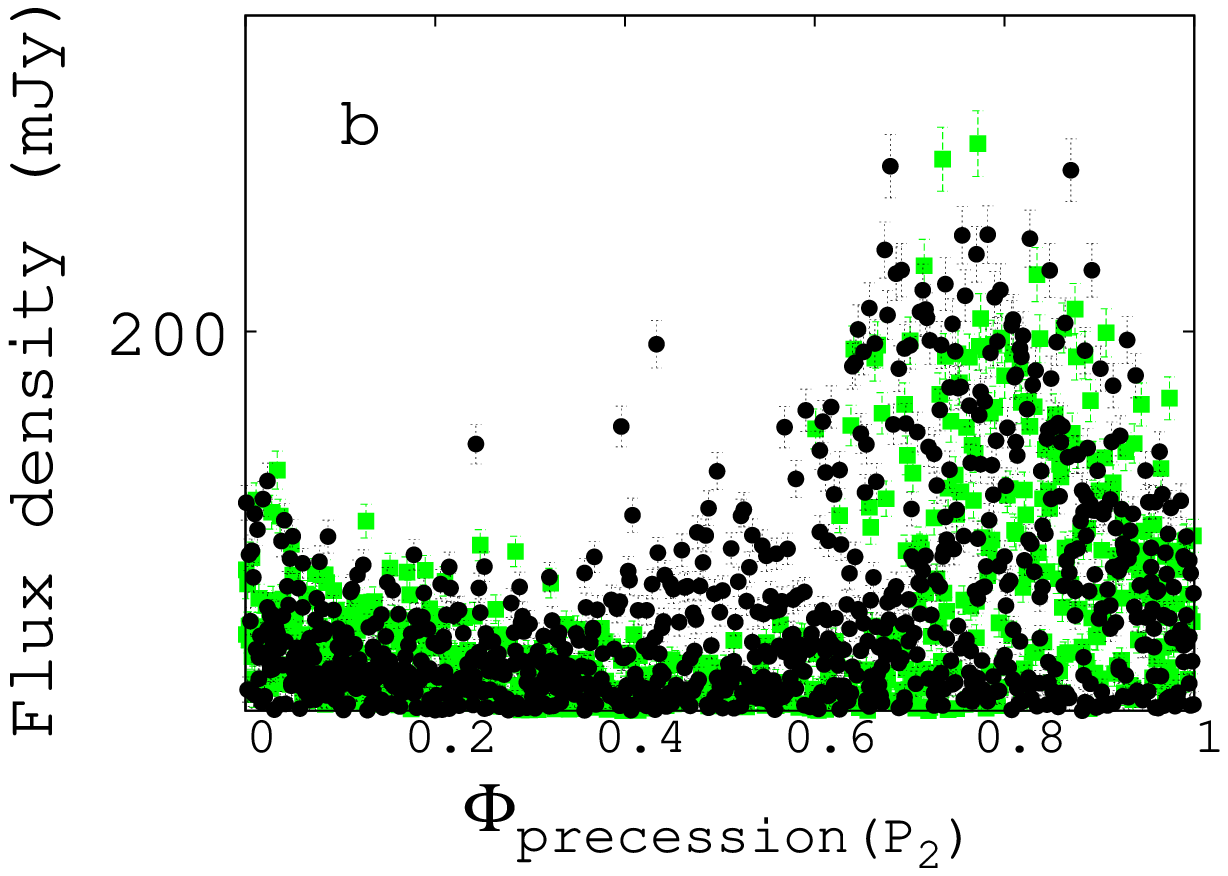}\\
   \includegraphics[width=.22\textwidth, height=.25\textwidth]{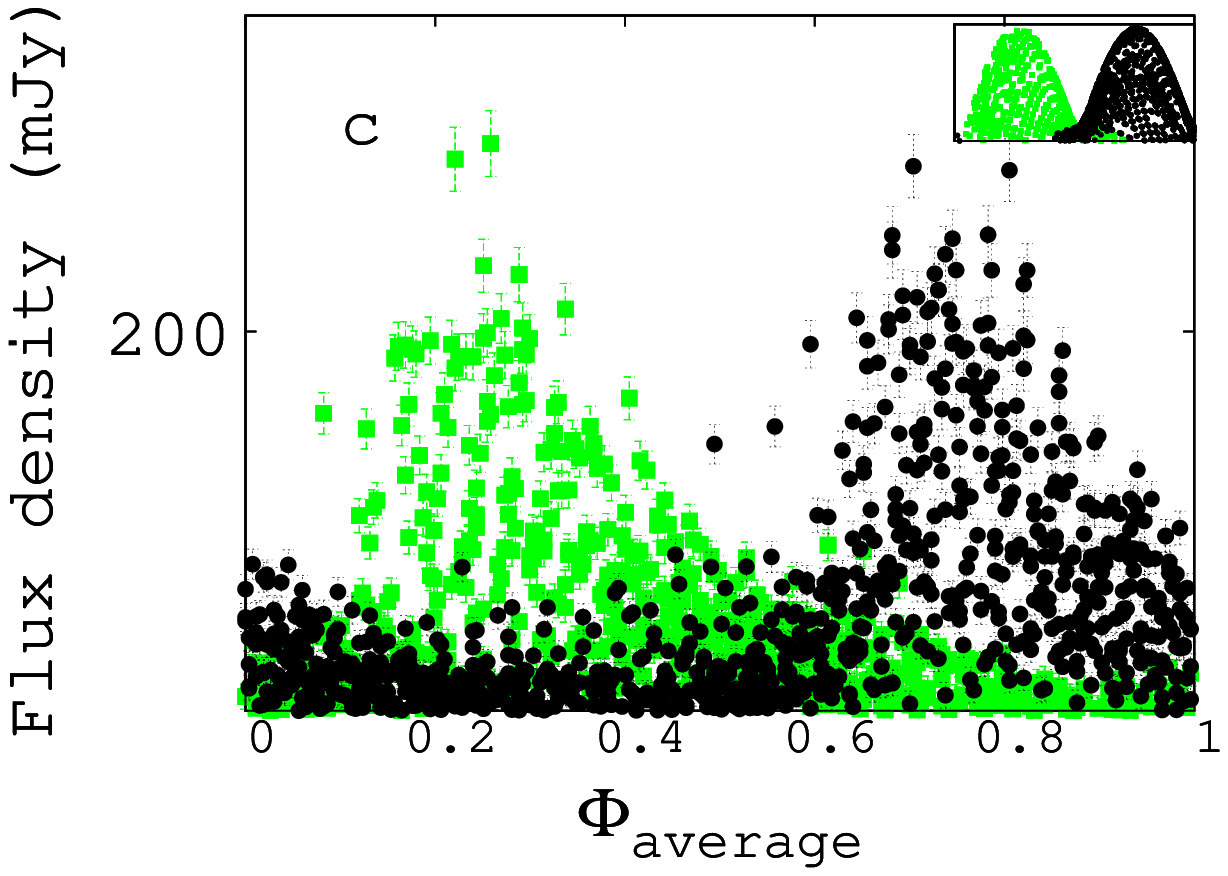}
   \includegraphics[width=.22\textwidth, height=.25\textwidth]{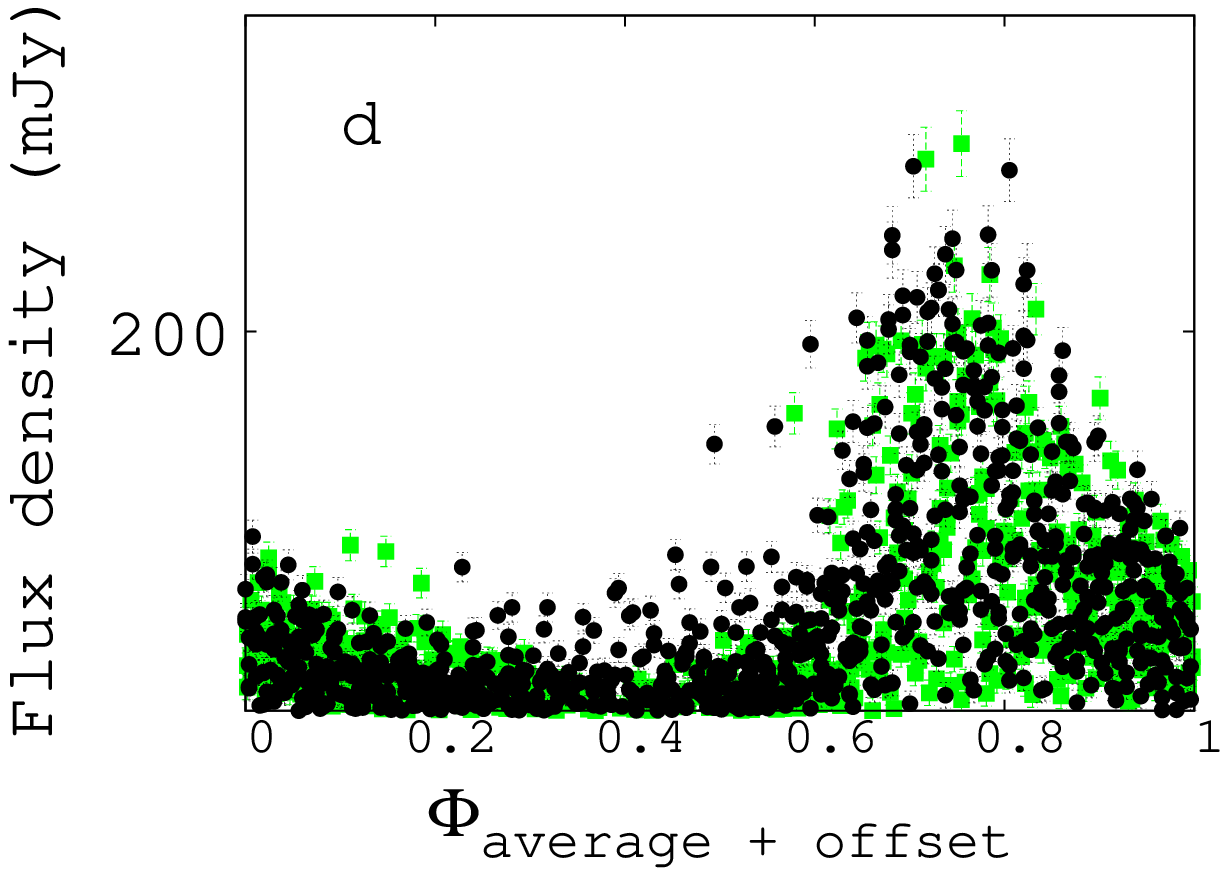}\\
    \includegraphics[scale=0.25]{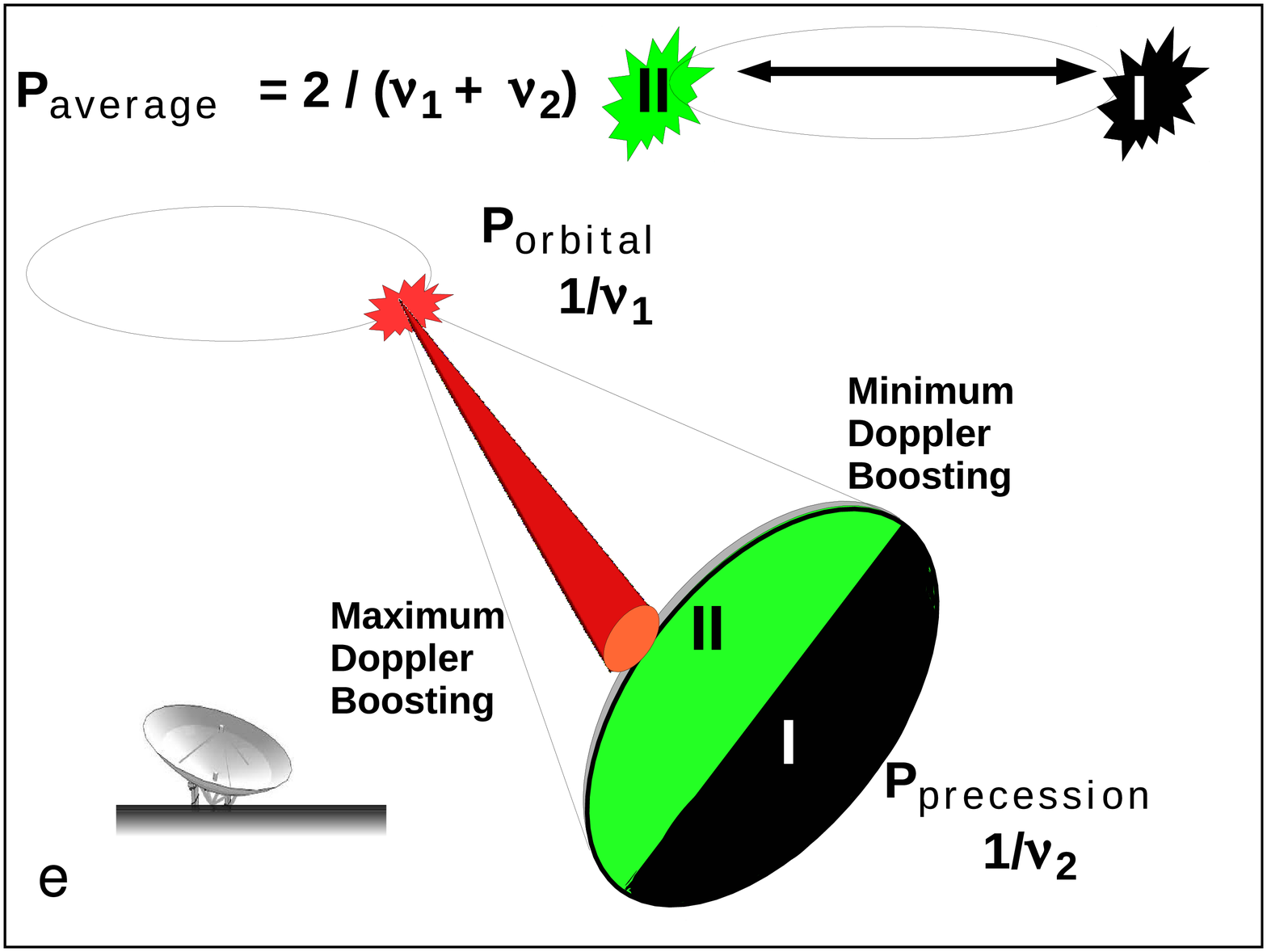}
  \caption{ 
	  a: Radio light curves (data averaged over 1 d) vs  $\Phi_{\rm orbit}$, with
	  $\Phi_{\rm orbit}$  related to 
    ${(t-t_0)\over P_1}$,
   with  $P_1$=26.49 d and 
     $t_0$=JD2443366.775  \citep{gregory02},
before (black)  and after (green) 
(2400000.5 + 50841)JD (see Appendix).
	  b: Same as ``4\,a'' but for  $\Phi_{\rm precession}$, related to $P_2$.
c: Radio light curves  vs $\Phi_{\rm average}$ for  $P_{\rm average}=26.70$ d.
 The small window shows the  simulated data (Eq. 2) identically folded.
d: Radio light curve  vs $\Phi_{\rm average}$ 
   with  $t_0$ changed by $\Delta t$ ($\Delta t=0$ before (2400000.5 + 50841)JD and
    $\Delta t=13.25$ d after it, see Appendix).
e: Sketch of the precessing jet in \lsi (out of scale).}
  \label{Fig4}
\end{figure}

\section{Conclusions and Discussion}
Our timing analysis and simulations give the following results:
\begin{enumerate}
\item
The timing analysis of 6.7~years of GBI data of \lsi results in two
frequencies
 $\nu_1=\unit[0.03775]{d^{-1}}$ ($P_1=\unit[26.49 \pm 0.07]{d}$) and
    $\nu_2=\unit[0.03715]{d^{-1}}$ ($P_2=\unit[26.92 \pm 0.07]{d}$).
The aim of our research, to obtain a better determination of the precessional period,
indicated by the astrometry to be  of 27--28 d, has therefore been reached. 
The period exists and it
is sligthly above the orbital period. 
\item
An additional and  totally unexpected result of our research is that    
    these two  periodicities give rise to  
  a  $P_{\rm average} =    \frac{2}{\nu_1 + \nu_2} = \frac{2} {0.03775 +0.03715}=
  \unit[26.70 \pm 0.05]{d}$,  modulated with 
  $P_{\rm beat}={1\over\nu_1-\nu_2}={1\over {0.03775 -0.03715}}= \unit[1667
  \pm 393]{d}$. In other words, the long-term periodicity 
  is equal to the beat of  $P_1$ and $P_2$, and  $P_{\rm beat}$  modulates a new period, $P_{\rm average}$. 
\item
We have shown that the  sawtooth function  derived in the past  \citep{gregory99}
by comparing observed and predicted (for $P=P_1$) outburst time
compares $P_1$ to  $P_{\rm average}$.
Our result therefore confirms  the controversial value of  $26.69\pm0.02$d obtained in the past by \citet{ray97}
for the periodicity of the outburst.
\item
$P_{\rm average}$ is only an apparent periodicity, a result of the beat of $P_1$ and $P_2$.
When the GBI  data are folded with
$P_{\rm average}$ the data, instead of clustering at one specific phase, 
cluster at two phases
separated by about 0.5 (13.25 d, i.e., $P_1/2$) depending 
on whether  the data are before or after  
the minimum of the long-term modulation. 
We have  shown that  the beat of $P_1$ and $P_2$
reproduces the same double clustering as the  GBI data.  
We find that  the time in the minimum of the long-term modulation when the phase jump
occurs is the time when the relative delay between the two functions
$f(P_1)$ and $f(P_2)$ have reached the maximum delay of $P_2/2$.
In a physical scenario this corresponds, as discussed below, to the point where the ejection has travelled half a 
 precession cone and  turns  on the other half of the precession cone.
\item
	$P_{\rm average}$ and $P_{\rm beat}$ are related to each other. They are both produced by  the beat between $P_1$ and $P_2$,
	the two real periodicities.
	$P_{\rm beat}$, the long-term periodicity, is like  $P_{\rm average}$,  just an apparent
	periodicity.
\item
$P_1$ and the  long-term modulation alone, cannot reproduce the observed periodogram
(i.e., $P_2$).
It would be possible to  reproduce the observed periodogram only by adding a  sawtooth 
function. In this scenario, the sawtooth function must  be produced
by a physical  process 
continuously changing $P_1$ to $P_{\rm average}$
 until a shift in orbital phase  of   +6 d/26.49 d is reached. Then 
another process would suddenly   shift the outburst from  +6 d/26.49 d  to 
 -7 d/26.49 d in orbital phase,
but without changing   the amplitude of the outburst (remaining in the minimum).
Instead,  we have shown that  using directly the two 
 periods $P_1$ and $P_2$ that we found in the periodogram of the GBI data, 
one naturally explains the two apparent periods $P_{\rm beat}$ and  $P_{\rm average}$, 
and the phase jump in  $P_{\rm average}$.  

\item
Implications of our results for variations at  other wavelengths  
indicate  the precessing relativistic jet as the agent responsible for the observed   H$\alpha$ variations,
equal to the variations that we observe  in  radio emission of the jet. 
\end{enumerate}

In conclusion, \lsi seems to be  one more case  in astronomy of a ``beat'',
i.e., a phenomenon occurring when two physical processes create 
stable variations  of  nearly equal frequencies.  The very small difference in frequency
creates a long-term variation of period 1/($\nu_1-\nu_2$). The first astronomical
case was that of a class  of Cepheids, afterwards called  
beat Cepheids  \citep{oosterhoff57}.  

The two periods in \lsi
 are  the precession $P_2$ of an  ejection having  orbital occurrence $P_1$.
Following the results of the radio spectral index analysis by  \citet{massikaufman09},
the steady ejection of relativistic electrons associated to the compact  object in \lsi 
increases until it terminates in a transient jet twice along the orbit.
As quoted in the introduction, this agrees well with the \citet{bondihoyle44}  accretion in an eccentric orbit
that  predicts two events in the system \lsp: one around periastron, and the second shifted 
towards apastron   \citep{taylor92, martiparedes95, boschramon06, romero07}.
However, the ejected relativistic electrons around periastron suffer
strong inverse Compton losses because they are exposed to stellar photons,
and only  at the second accretion/ejection peak,
occuring  rather displaced from periastron,
 the relativistic electrons survive  inverse Compton  losses and
 produce the large observed radio outburst with period  $P_1=26.49$\,d \citep{boschramon06}.
 This   ejection periodically changes direction \citep{massi12}   with
 a period  established in the present paper of 26.92\,d.
The radio maximum of the long-term modulation results  when the ejection 
occurs at the smallest  angle with
respect to the line of sight and the  Doppler boosting is largest 
\citep{kaufman02}.
The  radio minimum results when  the    ejection 
occurs at large angles with respect to the line of sight.
The point where the ejection has travelled half a 
 precession cone 
and  turns  on the other half of the precession cone
gives rise to the jump in phase for the apparent $P_{\rm average}$.

This research has brought up many questions in need of further investigation.
Future work needs to be done to establish the 
radio  vs H$\alpha$  relationship, the  
possible physical processes under the observed precession, 
and the Doppler boosting effects. 
Future observations are needed to 
estimate  the angle of the precession cone; 
an estimate of this  angle could result from comparing VLBA
 observations at the same orbital phase but interlapsed     (1667/2)d,
that is  by about 31  orbital cycles. From the different position angle of
the radio structures one could infer the aperture of the cone. 
Of course the two images interlapsed by 31 cycles should be done with high precision at the same 
orbital phase. Compare in Fig.~1 in \citet{massi12} image A (or image B) with  image I (or J)
 taken only one cycle later than A (B), but at a sligtly differente phase $\Delta
\Phi =0.016$. The small phase difference already causes differences in the images.
In \citet{albert08} it is discussed that two images taken 10 cycles apart 
and at the same orbital phase
 $\Phi=0.62$ are higly similar. Of course 10 cycles are still far from the requested 31 cycles.
However, the slightly different position angle between the two 
images in Fig.\,1 of those authors 
indicates that this comparison could be  a good investigative tool
to estimate  the precession angle in  future observations.

\acknowledgements
We would like to thank the anonymous referee for helpful comments.
We thank  Lisa Zimmermann and J\"urgen Neidh\"ofer for
reading the manuscript and for the several interesting discussions.
The Green Bank Interferometer is a facility of the National
Science Foundation operated by the NRAO in support of NASA High Energy
Astrophysics programs. 
\bibliographystyle{aa}

\begin{appendix}
\section{Beat and sawtooth function}

Let us assume two sine functions $f_1$ and $f_2$ with
period $P_1 = 26.49$\,d
and $P_2 = 26.92$\,d. 
Let us start during the radio maximum, i.e., with zero timing residual between
the peaks of $f_1$ and $f_2$.
In Fig.~4\,e a zero timing residual corresponds to an  ejection with 
the smallest angle with respect to the
line of sight, that is with the strongest   Doppler boosting.
Since $P_1$ is shorter than $P_2$,  each cycle $f(P_2)$ has a delay of
$P_2 - P_1 = \unit[0.43]{d}$.
The period  formed by the beat $P_{\rm average}$ 
has a  timing residual $\tau$ between its peak  and that of $f(P_1)$
half of that delay, i.e., $\unit[0.43/2=0.21]{d}$. 
This agrees well with   
 $\tau=0.008 \times P_1= 0.21$\,d of Eq. 4.

$f(P_{\rm average})$ is always between $f_1$ and $f_2$ and its peaks
have a regular distance of $P_{\rm average}= 26.70$~d (Fig.~3\,d).
However,  at the minimum  the accumulated delay of $f(P_2)$ with
respect  to $f(P_1)$  is almost $P_2/2$. That is, as shown in Fig.~3\,c, the 
 peak of $f(P_2)$ is nearly equidistant from the two
peaks of $f(P_1)$, the peak about 13 d before 
and the peak  about 13 d after it.
The beat with the first peak gives rise to a peak of $f(P_{\rm average})$ at 
about 827.8 d, whereas the beat with the second peak
gives rise to  a peak of $f(P_{\rm average})$ at 
841.05 d. The difference between these  two consecutive peaks of $f(P_{\rm average})$
is 13.25 d. This corresponds to  about 0.5 in phase, 
when the data are plotted in phase for $P_{\rm average}$ (Fig.~4\,c),
and also  corresponds to the  jump of about 13 d in the sawtooth function
observed  by \citet{gregory99}.
After that point,  $f(P_2)$ is preceeding  $f(P_1)$. 
In the physical scenario of Fig.~4\,e this jump,
or reset of the phase of $P_{\rm average}$,  corresponds to the 
 point where the ejection has travelled half a 
 precession cone 
and  turns  onto the other half of the precession cone.
Figure 3\,b shows  
$\tau$ vs time, resulting from this  simple analysis based on
the sine functions of Fig.~3; 
the resulting slope of the function is indeed 0.008, as in
Fig.~3\,a.
\end{appendix}
\end{document}